\documentclass[12pt,preprint]{aastex}
\slugcomment{Version: 11 March 2009, 11:00 EDT}
\shorttitle{Chandra COSMOS Survey}
\shortauthors{Elvis et al.}

\begin{document}
%
\title{The Chandra COSMOS Survey, I:\\ Overview and Point Source Catalog}
%
\author{Martin Elvis\altaffilmark{1}, Francesca
  Civano\altaffilmark{1}, Cristian Vignali\altaffilmark{2,3},
  Simonetta Puccetti\altaffilmark{4,5}, Fabrizio
  Fiore\altaffilmark{5}, Nico Cappelluti\altaffilmark{6},
  T. L. Aldcroft\altaffilmark{1}, Antonella Fruscione\altaffilmark{1},
  G. Zamorani\altaffilmark{3}, Andrea Comastri\altaffilmark{3},
  Marcella Brusa\altaffilmark{6,7}, Roberto Gilli\altaffilmark{3},
  Takamitsu Miyaji\altaffilmark{8,9}, Francesco
  Damiani\altaffilmark{10}, Anton Koekemoer\altaffilmark{11}, Alexis
  Finoguenov\altaffilmark{6,7}, Hermann Brunner\altaffilmark{6},
  C.M. Urry\altaffilmark{12}, John Silverman\altaffilmark{13},
  Vincenzo Mainieri\altaffilmark{14}, Guenther
  Hasinger\altaffilmark{6,15}, Richard Griffiths\altaffilmark{16},
  Marcella Carollo\altaffilmark{13}, Heng Hao\altaffilmark{1}, Luigi
  Guzzo\altaffilmark{17}, Andrew Blain\altaffilmark{18}, Daniela
  Calzetti\altaffilmark{19}, C. Carilli\altaffilmark{20}, Peter
  Capak\altaffilmark{21}, Stefano Ettori\altaffilmark{3}, Giuseppina
  Fabbiano\altaffilmark{1}, Chris Impey\altaffilmark{22}, Simon
  Lilly\altaffilmark{13}, Bahram Mobasher\altaffilmark{23}, Michael
  Rich\altaffilmark{24}, Mara Salvato\altaffilmark{18},
  D.B. Sanders\altaffilmark{25}, Eva Schinnerer\altaffilmark{26},
  N. Scoville \altaffilmark{18}, Patrick Shopbell\altaffilmark{18},
  James E. Taylor\altaffilmark{27}, Yoshiaki Taniguchi\altaffilmark{28},
  Marta Volonteri\altaffilmark{29}}
\altaffiltext{1}{Harvard-Smithsonian Center for Astrophysics, 60
  Garden St., Cambridge, MA 02138 USA}
\altaffiltext{3}{INAF$-$Osservatorio Astronomico di Bologna, Via
  Ranzani 1, I--40127 Bologna, Italy}
\altaffiltext{2}{Dipartimento di Astronomia, Universit\`a degli Studi
  di Bologna, Via Ranzani 1, I--40127 Bologna, Italy}
\altaffiltext{4}{ASI Science Data Center, via Galileo Galilei, 00044
  Frascati Italy}
\altaffiltext{5}{INAF$-$Osservatorio astronomico di Roma, Via Frascati
  33, 00040 Monteporzio Catone, Italy} 
\altaffiltext{6}{Max-Planck-Institute f\"ur Extraterrestrische Physik,
  Postfach 1312, 85741, Garching bei M\"unchen, Germany}
\altaffiltext{7}{University of Maryland, Baltimore County, 1000
 Hilltop Circle,  Baltimore, MD 21250, USA}
\altaffiltext{8}{Instituto de Astronom\'ia, Universidad Nacional Aut\'onoma
 de M\'exico, Ensenada, M\'exico (mailing address: PO Box 439027, San Ysidro,
 CA, 92143-9027, USA)}
\altaffiltext{9}{Center for Astrophysics and Space Sciences,
 University of California San Diego, Code 0424, 9500 Gilman Drive,
 La Jolla, CA 92093, USA}
\altaffiltext{10}{INAF - Osservatorio Astronomico di Palermo, Piazza
  del Parlamento 1, I-90134 Palermo, Italy}
\altaffiltext{11}{Space Telescope Science Institute, 3700 San Martin
  Drive, Baltimore, MD 21218 USA}
\altaffiltext{12}{Department of Physics and Yale Center for Astronomy
  \& Astrophysics, Yale University, P.O. Box 208121, New Haven, CT
  06520-8121, USA} 
\altaffiltext{13}{Department of Physics, Eidgenössische Technische
  Hochschule-Zurich, CH-8093 Zurich, Switzerland}
\altaffiltext{14}{ESO, Karl-Schwarschild-Strasse 2, D--85748 Garching, Germany}
\altaffiltext{15}{Max-Planck-Institute f\"ur Plasmaphysik,
  Boltzmannstrasse 2, D-85748 Garching bei M\"{u}nchen}
\altaffiltext{16}{Department of Physics, Carnegie Mellon University,
  5000 Forbes Avenue, Pittsburgh, PA 15213 USA}
\altaffiltext{17}{INAF$-$Osservatorio Astronomico di Brera, via Bianchi
  46, 23807 Merate, Italy} 
\altaffiltext{18}{California Institute of Technology, MC 105-24, 1200
  East California Boulevard, Pasadena, CA 91125 UA}
\altaffiltext{19}{Department of Astronomy, University of Massachusetts,
  Amherst, MA 01003 USA}
\altaffiltext{20}{National Radio Astronomy Observatory, PO Box O,
  Socorro NM 87801 USA}
\altaffiltext{21}{Spitzer Science Center, 314-6 Caltech, Pasadena, CA
  91125 USA}
\altaffiltext{22}{Steward Observatory, University of Arizona, 933
  North Cherry Avenue, Tucson, AZ 85721 USA}
\altaffiltext{23}{Department of Physics and Astronomy, University of
  California, Riverside, CA 92521 USA}
\altaffiltext{24}{Department of Physics and Astronomy, University of
  California, Los Angeles, CA 90095 USA}
\altaffiltext{25}{Institute for Astronomy, University of Hawaii, 2680
  Woodlawn Dr., Honolulu, HI 96822 USA}
\altaffiltext{26}{Max-Planck-Institut f\"{u}r Astronomie, K\"{o}nigstuhl 17,
  Heidelberg D-69117, Germany} 
\altaffiltext{27}{Department of Physics and Astronomy, University of
  Waterloo, Waterloo, Ontario, N2L 3G1, Canada}
\altaffiltext{28}{Research Center for Space and Cosmic Evolution,
  Ehime University, Bunkyo-cho 2-5, Matsuyama 790-8577, Japan}
\altaffiltext{29}{Department of Astronomy, University of Michigan, Ann
  Arbor, MI 48109 USA}
%
\begin{abstract}
The {\em Chandra} COSMOS Survey (C-COSMOS) is a large, 1.8~Ms, {\em
Chandra} program that has imaged the central 0.5~sq.deg of the COSMOS
field (centered at 10$^h$, +02$^{o}$) with an effective exposure of
$\sim$160~ksec, and an outer 0.4~sq.deg. area with an effective
exposure of $\sim$80~ksec. The limiting source detection depths are
1.9$\times$10$^{-16}$~erg~cm$^{-2}$~s$^{-1}$ in the Soft (0.5--2~keV)
band, 7.3$\times$10$^{-16}$~erg~cm$^{-2}$~s$^{-1}$ in the Hard
(2--10~keV) band, and 5.7$\times$10$^{-16}$~erg~cm$^{-2}$~s$^{-1}$ in
the Full (0.5--10~keV) band.
Here we describe the strategy, design and execution of the C-COSMOS
survey, and present the catalog of 1761 point sources detected at a
probability of being spurious of $<$2$\times$10$^{-5}$ (1655 in the Full,
1340 in the Soft, and 1017 in the Hard bands). 
By using a grid of 36 heavily ($\sim$50\%) overlapping pointing
positions with the ACIS-I imager, a remarkably uniform ($\pm$12\%)
exposure across the inner 0.5~sq.deg field was obtained, leading to a
sharply defined lower flux limit.
The widely different PSFs obtained in each exposure at each point in the field
required a novel source detection method, because of the overlapping tiling
strategy, which is described in a companion paper. 
This method produced reliable sources down to a 7--12 counts, as verified by the
resulting logN-logS curve, with sub-arcsecond positions, enabling optical and
infrared identifications of virtually all sources, as reported in a second
companion paper. 
The full catalog is described here in detail and is available on-line.

\end{abstract}
%

\section{Introduction}

The co-evolution of galaxies and quasars or active galactic nuclei (AGNs) has
been vigorously pursued both observationally and theoretically for a decade,
ever since the discovery that the mass of the central black hole is tightly
correlated both with the luminosity (Magorrian et al. 1998; Marconi \& Hunt
2003) and the velocity dispersion of the spheroid (M$_{BH}$-$\sigma$ relation;
Ferrarese \& Merrit 2000; Gebhardt et al. 2000; Tremaine et al. 2002).  Tackling
this large subject requires the study of both galaxies and AGNs back to the
epoch when both were growing rapidly, i.e. z$\sim$1-3, requiring deep
observations across many wavelengths, from radio through the infrared, optical
and ultraviolet, to the X-rays. At the same time, the wide range of cosmic
density and the rapid changes in this large scale structure (LSS) require wide
field observations that sample the Universe at close to their true fractions.

The Cosmic Evolution Survey (COSMOS, Scoville et al. 2007a) is a deep and wide
extragalactic survey designed to have sufficient area to overcome most cosmic
variance, which requires sampling regions some 50~Mpc on a side (Fig.1; Scoville
et al. 2007a), and with sufficient depth to sample the z = 1 - 3 galaxy and AGN
population. The contiguous 2~sq.deg COSMOS field samples a volume of
$\sim$6$\times$10$^{6}$~Mpc$^3$ at $z=0.5-1$ (Wright 2006). This is $\sim$10\%
of the volume imaged by the Sloan Digital Sky Survey (SDSS) in the local
(z$<$0.1) universe (5.7$\times$10$^7$Mpc$^3$, 8000~sq.deg, DR5\footnotemark).
\footnotetext{URL: http://www.sdss.org/dr5/}
COSMOS is a region of low, uniform, Galactic obscuration
(E(B-V)$\simeq$0.02~mag, N$_H$ (2.7$\times$10$^{20}$cm$^{-2}$, Dickey \& Lockman
1990). COSMOS is likely to be the largest survey of this type for the next
decade.

\begin{figure}[t]
\includegraphics[width=\textwidth]{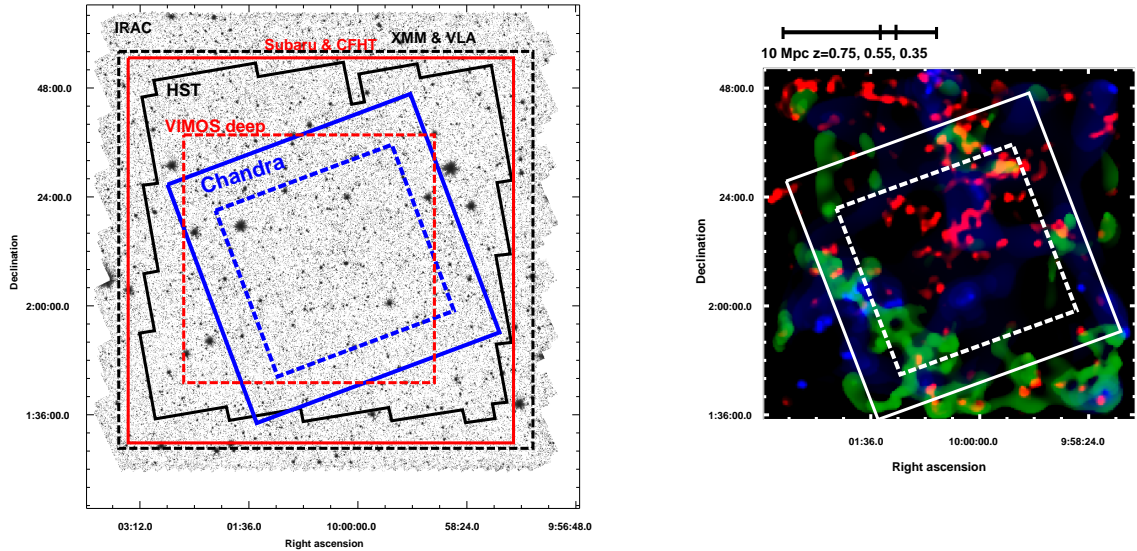}
\vspace{-6mm}
\caption{\small {\em Left:} Map of the COSMOS field showing the coverage at
    various wavelengths: the IRAC 3.6$\mu$m mosaic is the background image; Blue
    solid={\em Chandra}, blue dashed={\em Chandra} deep; black polygon = HST; red solid =
    Subaru, CFHT, zCOSMOS bright; red dashed = zCOSMOS deep; black dashed = XMM
    and VLA. The {\em Spitzer} MIPS observations cover an area 2 times larger.
    {\em Right:} Large scale structure seen in galaxy distributions in the
    COSMOS field (Scoville et al. 2007c), ranging in size from 1~-~20~Mpc, based
    on photo-z's accurate to $\sim$1\%. Blue is centered on z=0.35, Green on
    z=0.55, and Red on z=0.75, each with $\Delta$z=0.05.  The C-COSMOS field
    outline is shown as the white tilted square, with the dashed line
    delineating the high exposure area as in the left panel. A scale showing
    10~Mpc at the three redshifts is shown at the top. In both panels North is
    up, East is to the left.}
\label{cosmos}
\end{figure}

The location of the COSMOS area near the equator (10$^h$, +02$^{\circ}$) allows
all major and future facilities\footnote{Except for those in Antartica.}
(notably EVLA, ALMA, and SKA) to target this region down to faint limits
(Scoville et al. 2007a). Space-based imaging has been undertaken in the F840W
($\sim i$-band) with {\em Hubble Space Telescope} (HST, Scoville et al. 2007b),
in the 3.5~$\mu$m-70$\mu$m infrared using {\em Spitzer} IRAC and MIPS (Sanders
et al. 2007), in the UV using GALEX (Zamojski et al.  2007), and in 0.5-10~keV
X-rays with XMM (Hasinger et al. 2007, Cappelluti et al. 2007). Ground-based
imaging spans the radio (1400~MHz VLA, Schinnerer et al. 2007), the near-IR with
CTIO and KPNO (Capak et al. 2007) and CFHT (McCracken et al. 2009, in
preparation), the optical to AB$\sim$26-27 with {\em Subaru} in 21 bands
(Taniguchi et al. 2007). Finally, large dedicated ground-based spectroscopy
programs in the optical with Magellan/IMACS (Trump et al. 2007), and VLT/VIMOS
(Lilly et al. 2008) are well underway.

This wealth of data has resulted in an initial 15-band photometric catalog of
$\sim$10$^6$ objects (Capak et al. 2007) from which photometric redshifts good
to $<$3\% for z$<$1.2 and $r<$24 have been derived (Mobasher et
al. 2007). Recently, more photometric bands have been added, resulting in
improved photo-z's for the galaxy population accurate to $\Delta$z/(1+z)$<$1\%
(Ilbert et al. 2009) and to $\Delta$z/(1+z)$\sim$2\% for the AGN population
(Salvato et al. 2009).

We have undertaken the {\em Chandra}-COSMOS survey (C-COSMOS) to cover the
central 0.9~sq.deg region of the COSMOS field (Fig.\ref{cosmos}, left),
containing a wide range of cosmic overdensity (Fig.\ref{cosmos}, right), with
the ACIS-I CCD imager (Garmire et al. 2003) on board the {\em Chandra X-ray
  Observatory} (Weisskopf et al. 2002). The survey took 1.8~Msec of {\em
  Chandra} observing time ($\sim$21~days) and was the largest guest observer
program approved in a single AO at the time it was undertaken (2006 November -
2007 June). C-COSMOS employed a series of 36 heavily overlapped ACIS-I 50~ksec
pointings to give an exposure of $\sim$160~ksec over the inner area to a depth
of $\sim$1.9$\times$10$^{-16}$erg~cm$^{-2}$s$^{-1}$ (0.5-2~keV), providing an
unprecedented combination of contiguous area and depth in the X-ray band.  This
overlapping tiling strategy gives highly uniform exposure, and so a well-defined
flux limit.

Several of the deepest COSMOS surveys are now concentrating on this same central
sub-field of COSMOS: the z-COSMOS Deep spectroscopic survey (to B$\sim$25, Lilly
et al. 2007), the deep VLA survey (6~$\mu$Jy rms, Schinnerer et al. 2009, in
preparation), and several millimeter and sub-millimeter surveys (MAMBO, Bertoldi
et al. 2007 and AzTEC, Scott et al. 2008).  GALEX has observed the central field
deeply (Zamojksi et al. 2007) and is currently monitoring this area. The
Ultra-VISTA survey will undertake a deep yJHK survey of the central 1.5~ sq.deg,
half of which will be surveyed to the unprecedented limits of $\sim$26~AB~mag
(Arnaboldi et al. 2007).

By going for large area rather than extreme depth, most of the C-COSMOS sources
are sufficiently bright to be detected in the rest of the pan-chromatic COSMOS
data set, allowing rapid identifications (Civano et al. 2009) and determination
of their multi-wavelength properties (e.g. Elvis et al. 2009, in
preparation). On the other hand, C-COSMOS is sufficiently deep that significant
numbers of normal and starburst galaxies with luminosity of
10$^{42}$~erg~s$^{-1}$ can be detected up to $z\sim$0.9, a redshift depth
comparable with that of the galaxy redshift surveys in the COSMOS field
(Taniguchi et al. 2007, Lilly et al. 2007).  Adding the {\em Chandra} coverage
to the COSMOS survey adds a valuable resource for the study of the co-evolution
of black holes and their host galaxies, of the SEDs of faint quasars and active
galactic nuclei, and the evolution of galaxies.

\begin{figure}
\begin{center}
\includegraphics[width=0.8\textwidth]{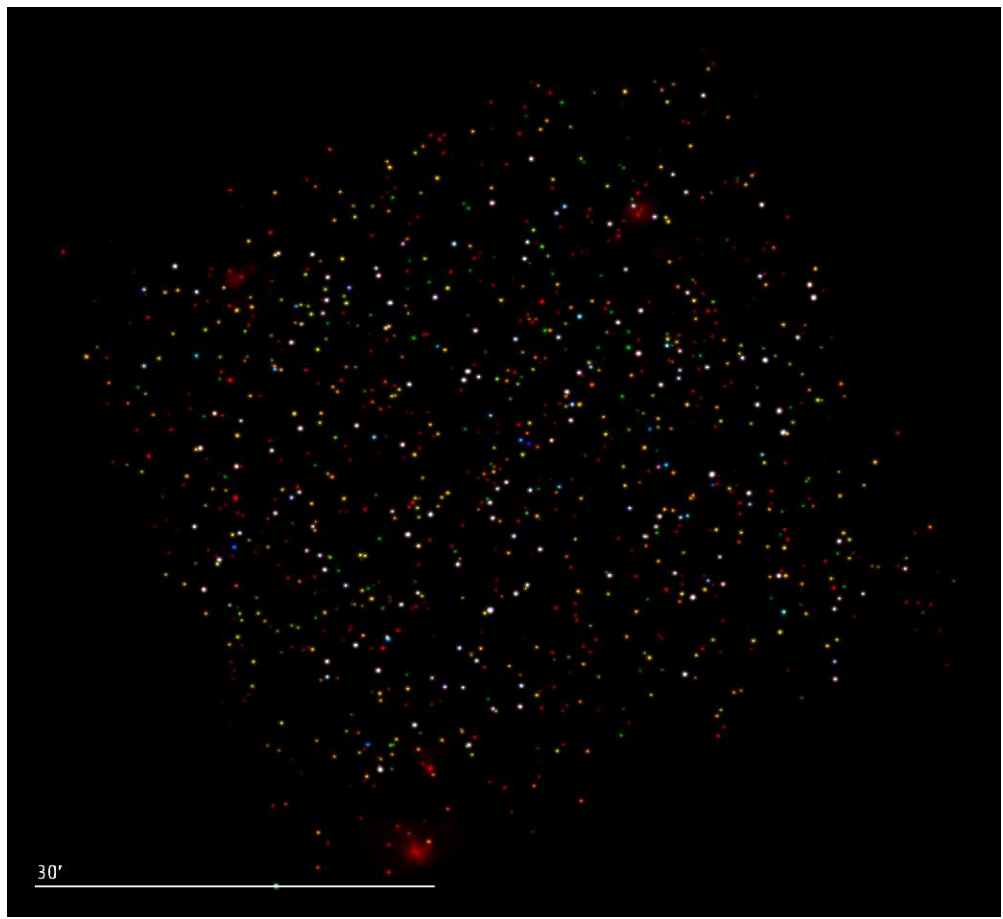}
\end{center}
\vspace{-4mm}
\caption{\small X-ray image of the Chandra COSMOS field, optimized to show point
  sources with a wide variety of X-ray colors. A scale showing 30~arcmin (the
  approximate diameter of the full moon) is shown for comparison.  North is
    at the top; East is to the left. The full angular resolution of {\em
    Chandra} is not well represented in this image as, in order to display the
  point sources clearly, the original image has been smoothed with a sharp
  gaussian with radius equal to 2.9$^{\prime\prime}$, and added to an image of
  the field smoothed with a wide gaussian with radius equal to
  4.4$^{\prime\prime}$.  X-ray 'colors' are mapped so that red is the 0.5-2~keV
  band, green is the 2-4.5~keV band, Blue is the 4.5-7~keV band, and each
  energy band was smoothed in the same way. Selected prominent clusters have
  been adaptively smoothed for display (red extended shapes).}
\label{ccosmos}
\end{figure}

The summed image of the entire C-COSMOS field is shown in Fig.\ref{ccosmos}
where colors have been mapped to X-ray bands. 

This is the first of three papers presenting the basic results of the C-COSMOS
survey over the whole field. Paper~I (this paper) reports on the strategy,
design and execution of the C-COSMOS survey, and present the catalog of 1761
point-like X-ray sources detected in C-COSMOS; Paper~II (Puccetti et al. 2009)
presents the details of the simulations carried out to optimize the source
detection method; Paper~III (Civano et al. 2009) presents the identification of
the X-ray sources with optical and infrared counterparts.
We conclude by listing the primary science objectives foreseen for the C-COSMOS
data. Papers on several of these topics are in preparation.

We assume a $\Lambda$CDM cosmology with H$_0$=70~km~s$^{-1}$,
$\Omega_m$=0.27, $\Omega_{vac}$=0.73.

\section{The Chandra COSMOS Strategy}

\begin{figure} 
\begin{center}
\includegraphics[width=0.6\textwidth]{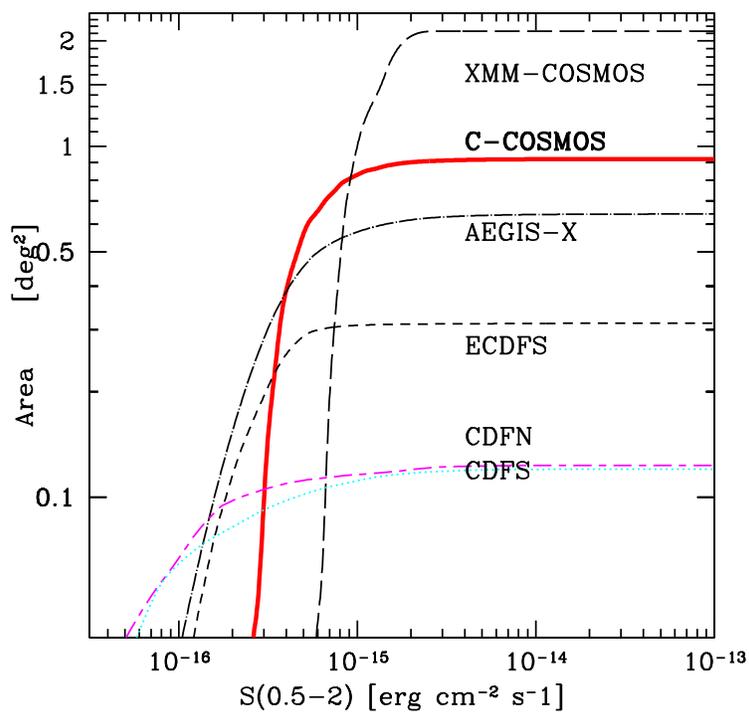}
\end{center}
\vspace{-0.5cm}
\caption{\small Area-flux curve for C-COSMOS (red solid line,
  0.5-2~keV).  The coverage of ECDFS (Lehmer et al. 2005; dashed line), 
  AEGIS-X (Laird et al. 2008; dash-dotted line), CDFN
  (Alexander et al. 2003; magenta short-long dashed line), CDFS (Luo et
  al. 2008; cyan dotted line) and XMM-COSMOS (Cappelluti et al. 2009;
  black dashed line) are shown for comparison.}
\label{AreaFluxCurve}
\end{figure}

For C-COSMOS we have developed a strategy that uses $\sim$50\% overlapping
tiling of the  16.9$\times$16.9~arcmin ACIS-I fields.  This tiling produces
a remarkably uniform sensitivity in the central part of the field, and a
well-defined flux limit with a sharp cut-off (Fig.\ref{AreaFluxCurve}; for
details on the generation of sensitivity maps see \S 7 in Paper II).  This
approach also ensures that the area with HPD$<$2$^{\prime\prime}$ is maximized,
so that the unique {\em Chandra} high resolution imaging (van~Speybroeck et
al. 2002) can be exploited fully, albeit with 1/4 of the exposure time. The good
{\em Chandra} point spread function (PSF) resolves sources 2$^{\prime\prime}$
apart over $\sim$0.7~sq.deg, corresponding to 8-16~kpc separations for
$z$=0.3-0.9, and locates point sources to $<$4~kpc at {\em any redshift}. Thus
close mergers can be resolved, and nuclear sources distinguished from
off-nuclear sources in galaxies (Ultra-luminous X-ray Sources, ULXs, Fabbiano
2006, Lehmer et al. 2006, Mainieri et al. 2009, in preparation).

Point source detection sensitivities were estimated for three standard
{\em Chandra} bands: Soft (S, 0.5-2~keV), Hard (H, 2-10~keV) and Full
(F, 0.5--10~keV). Due to the high background in the 7-10 keV energy
range\footnotemark, channels above 7~keV were not used for source
detection. (See \S 4.2.2 and Paper~II for details).
\footnotetext{URL: http://cxc.harvard.edu/contrib/maxim/bg/index.html\#spec}
The C-COSMOS flux limits in 3 bands are reported in Table~\ref{tabflux}, 
together with the XMM-COSMOS limits for comparison: C-COSMOS sensitivity is 
 three times below the corresponding flux limits for
the XMM-COSMOS survey (dashed line; Cappelluti et al. 2009), making them 
complementary surveys.

The achieved sensitivity-area curve\footnotemark
(Fig.~\ref{AreaFluxCurve}) has a sharp cut-off at low fluxes.

\footnotetext{This curve is remarkably close to the predictions
  from the proposal, reflecting the high accuracy with which the
  requested tiling was executed.}

\begin{table}[t]
\small
\caption{\small C-COSMOS flux limits and corresponding XMM-COSMOS flux limits.}
\begin{center}
\begin{tabular}{l c c r }
\hline
Band &C-COSMOS(lim)$^a$ & C-COSMOS(logN-logS)$^a$ & XMM-COSMOS$^a$ \\
\hline
Soft (0.5--2~keV) & 1.9 & 2.5 &  5 \\
Hard  (2--10~keV) & 7.3 & 16  & 25 \\
Full (0.5--10~keV)& 5.7 & --- & ---\\
\hline
\end{tabular}

$^a$ flux limits are reported in units of
10$^{-16}$erg~cm$^{-2}$s$^{-1}$, for bands up to 10~keV, but were
measured only up to 7~keV. (See text for details.)
\end{center}
\label{tabflux}
\end{table} 

The C-COSMOS Soft band flux limit corresponds to luminosities of (0.8,
4, 11)$\times 10^{41}$erg~s$^{-1}$ at $z$=(0.3, 0.6, 0.9)
respectively, while the Hard band flux limit corresponds to four times
higher luminosities. Both luminous elliptical galaxies and starbursts
often exceed these luminosities, and starburst galaxies are known to
become common (Hornschemeier et al. 2003) at these X-ray fluxes.

The low ACIS background enables stacking analysis, in which counts at
the positions of known classes of objects, e.g. subsets of the
thousands of galaxies with redshifts, are co-added to increase the
effective exposure time (Brusa et al. 2002; Hornschemeier et al.
2002; Brandt et al. 2001; Nandra et al. 2002; Fiore et al. 2008a,b).

\subsection{Design}

\begin{figure}[t]
\includegraphics[width=0.51\textwidth]{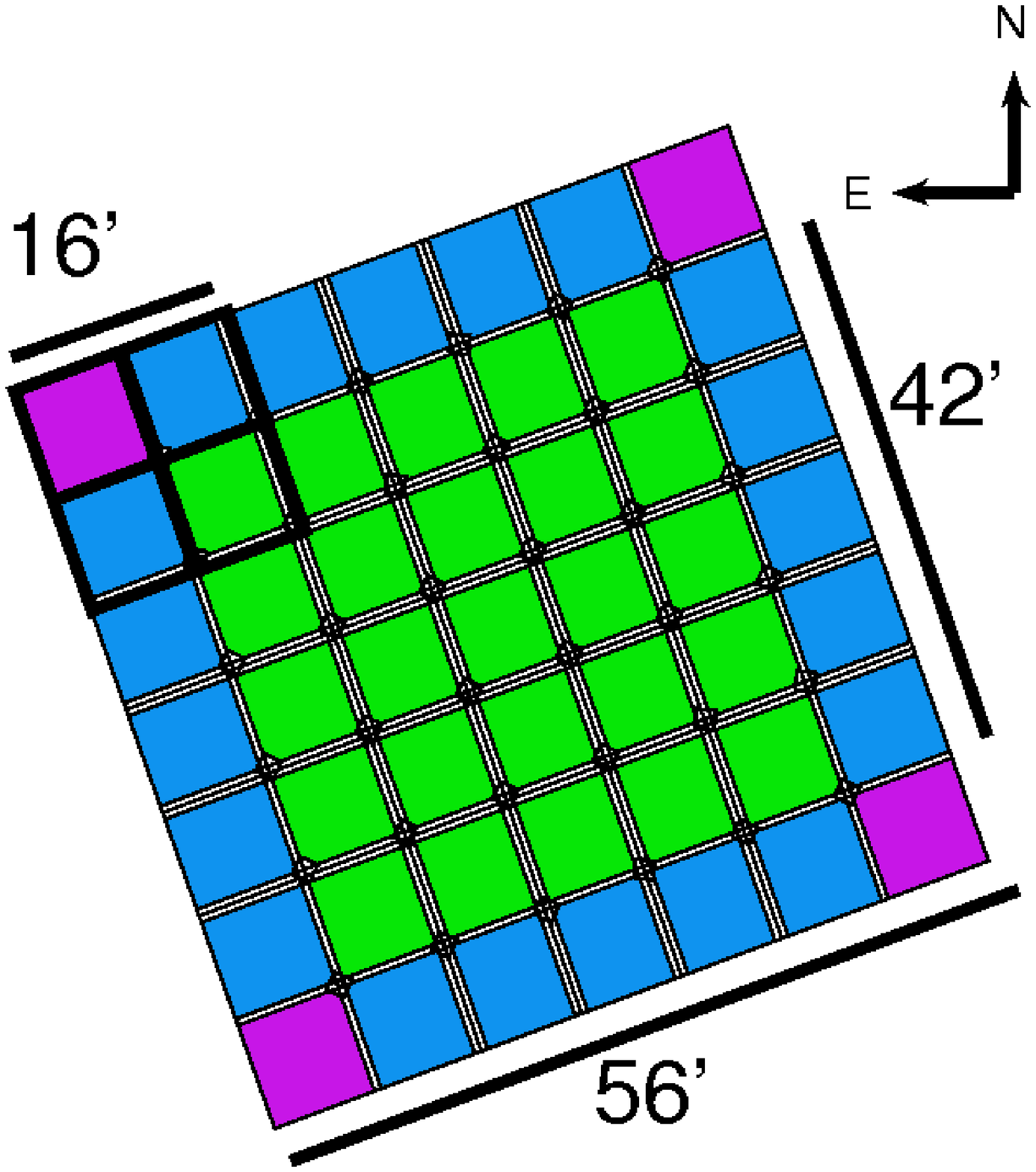}
\includegraphics[width=0.48\textwidth]{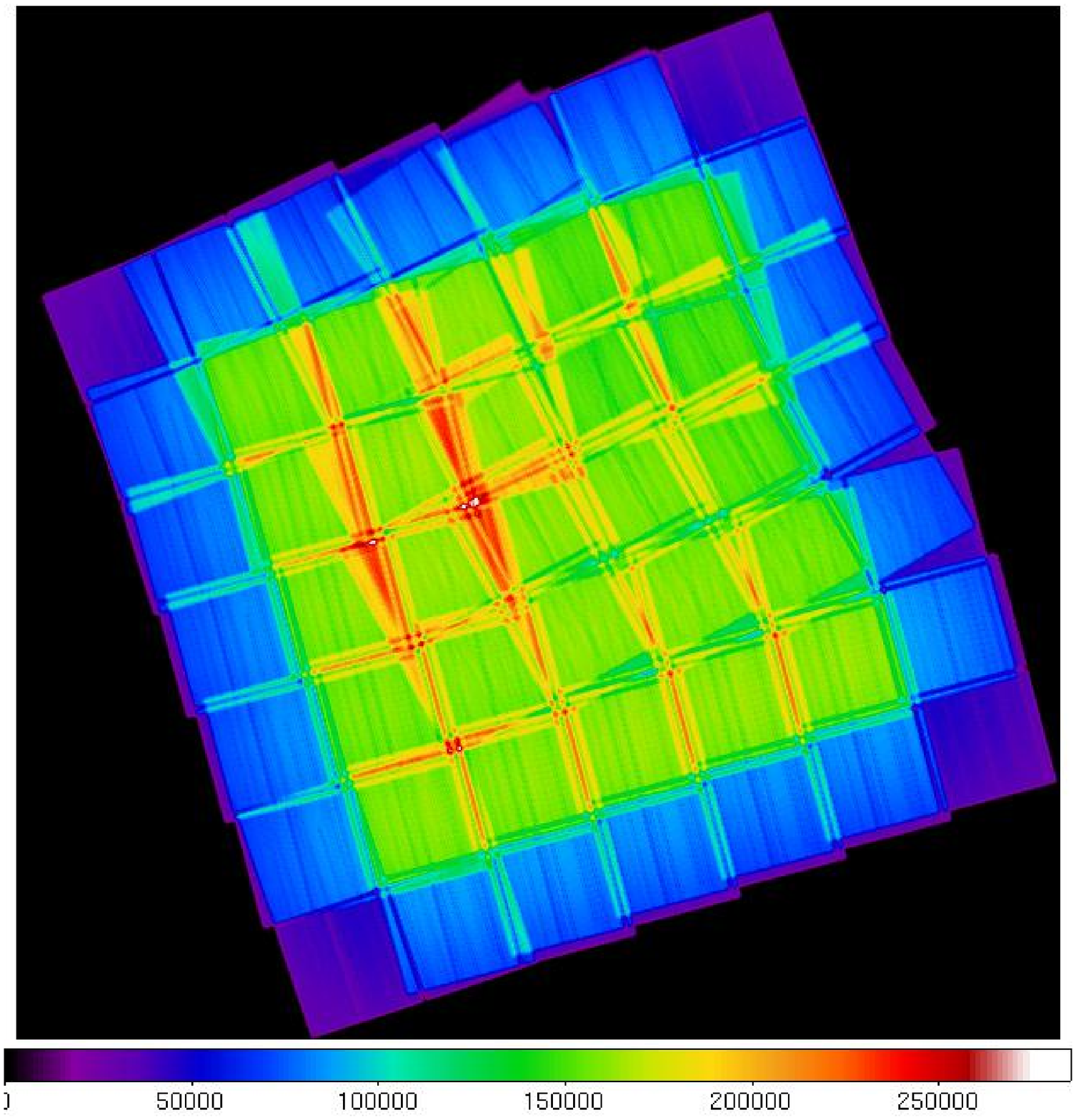}
\caption{\small {\em Left:} The 'as designed' C-COSMOS tiling for the 36 50~ksec
  pointings.  The thick black box (top left) represents one ACIS-I pointing, the
  thin boxes all the pointings. Different colors show areas with different
  number of overlapping pointings: green - 4 overlapping pointings; blue - 2
  overlapping pointings; purple - 1 pointing. The black bars show roughly the
  relative dimensions of one pointing ($\sim$16$\arcmin$), of the inner area
  with larger exposure ($\sim$42$\arcmin$), and of the total field
  ($\sim$56$\arcmin$). Raster point (see Table~\ref{tab:obs_summary} 1-1 lies at
  the top right (NE) and 1-6 lies at the top-left (NW).  {\em Right:} The 'as
  executed' exposure map for the C-COSMOS survey in the Soft band.  The color
  bar gives the achieved effective exposure in units of seconds.}
\label{tiling}
\end{figure}

The C-COSMOS tiling scheme (Fig. \ref{tiling}, left panel) covers the central
area of the COSMOS field in the most efficient manner that we could devise. A
6$\times$6 raster array of 36 ACIS-I pointings (one ACIS pointing field of view
is outlined in black in Fig. \ref{tiling}, {\em left}), each of 50~ksec nominal
exposure, were chosen. The center of the array (Table~\ref{corners}) is slightly
offset from the center of the COSMOS field to match the z-COSMOS deep field
(Lilly et al. 2007).

The value of the 8.0$^{\prime}$ offset between pointing centers was chosen to be
slightly less than the 8.3$^{\prime}$ size of an ACIS chip (Garmire et al.,
2003, Chandra Proposers' Observatory Guide, aka POG, 2007\footnotemark)
Table~6.1), so that chip gaps are not co-added to create small scale dips in the
effective exposure time.
\footnotetext{Chandra X-ray Center publication TD~403.00.010}

The inner part of the field was covered by four exposures, to give a total
nominal exposure of 200~ksec (effective exposure $\sim$160~ksec) over a
42$^{\prime}\times$42$^{\prime}$ area (0.5~sq.deg,green area in
Fig. \ref{tiling}).  The outer region has been covered by two observations (blue
area) and the four corners covered by 1 observation (purple area).  The corners
of the outer and inner regions are reported in Table \ref{corners} clockwise
from the top left.

Sources at a flux of $\sim$2.0$\times$10$^{-16}$erg~cm$^{-2}$s$^{-1}$
(0.5-2~keV) have a total of 5--10 summed counts in the four exposures,
ensuring a good detection, given the low {\em Chandra}/ACIS background
of $\sim$2~counts/200~ksec over a 2~arcsec radius circle (see \S
4.2.1).

The heavily overlapped tiling scheme produces a smooth exposure map
that is flat to 12\% in the central region (see Figure \ref{tiling},
right panel and \S 4.2.2).

\begin{center}
\begin{table}[t]
\footnotesize
\caption{Coordinates of the C-COSMOS field, center and corners of the outer and
  inner regions, clockwise from the NE (top left). }
\smallskip
\begin{center}
\begin{tabular}{c c }
\hline
RA  & Dec \\
\hline
\multicolumn{2}{l}{\em Center}\\
10$^h$~ 00$^m$~ 24$^s$&  +02$^{\circ}$~10$^{\prime}$~55$^{\prime\prime}$ \\
\hline
\multicolumn{2}{l}{\em Outer region}\\
10$^h$~ 02$^m$~45$^s$ &  +02$^{\circ}$~26$^{\prime}$~47$^{\prime\prime}$ \\  
09$^h$~ 59$^m$~11$^s$ &  +02$^{\circ}$~46$^{\prime}$~45$^{\prime\prime}$ \\  
09$^h$~ 57$^m$~54$^s$ &  +01$^{\circ}$~53$^{\prime}$~00$^{\prime\prime}$ \\  
10$^h$~ 01$^m$~23$^s$ &  +01$^{\circ}$~33$^{\prime}$~59$^{\prime\prime}$ \\
\hline
\multicolumn{2}{l}{\em Inner region}\\
10$^h$~ 02$^m$~05$^s$ &  +02$^{\circ}$~21$^{\prime}$~13$^{\prime\prime}$ \\ 
09$^h$~ 59$^m$~30$^s$ &  +02$^{\circ}$~35$^{\prime}$~47$^{\prime\prime}$ \\  
09$^h$~ 58$^m$~35$^s$ &  +01$^{\circ}$~59$^{\prime}$~19$^{\prime\prime}$ \\  
10$^h$~ 01$^m$~11$^s$ &  +01$^{\circ}$~44$^{\prime}$~37$^{\prime\prime}$ \\
\hline
\end{tabular}
\end{center}
\label{corners}
\end{table} 
\end{center}

\subsection{Comparison with Other Legacy Surveys}

{\em Chandra} observing time has been dedicated to several large legacy surveys:
CDF-S (Giacconi et al. 2002; Luo et al. 2008), CDF-N (Alexander et al. 2003),
ECDF-S (Lehmer et al. 2005), AEGIS-X (Nandra et al. 2005; Laird et al. 2008),
XBootes (Murray et al. 2005).  These surveys have different emphases in area and
depth, so we summarize the special features of C-COSMOS here.

Like all contiguous area surveys, C-COSMOS has significant advantages over
non-contiguous surveys (e.g. SEXSI, Harrison et al. 2003, Eckart et al. 2006;
ChaMP, Kim et al. 2007), because of the difficulty of getting deep
multi-wavelength coverage of non-contiguous fields.

C-COSMOS is neither the deepest (CDFN and CDFS) nor the widest
(XBootes) legacy {\em Chandra} survey.  A comparable sensitivity has
been reached in the somewhat smaller AEGIS field (dot-dashed line in
Fig.~\ref{AreaFluxCurve}; Laird et al. 2008).  C-COSMOS
differs from the other surveys by having the largest area at fluxes
$0.3-1\times 10^{-15}$erg~cm$^{-2}$s$^{-1}$, and a sharper low flux
limit cut-off in the area surveyed than most other recent X-ray
surveys. The single field CDF-S and CDF-N have notably shallower
roll-offs in their sensitivity curves (magenta and blue lines in 
Fig.~\ref{AreaFluxCurve}).

\begin{figure}
\begin{center}
\includegraphics[width=0.6\textwidth]{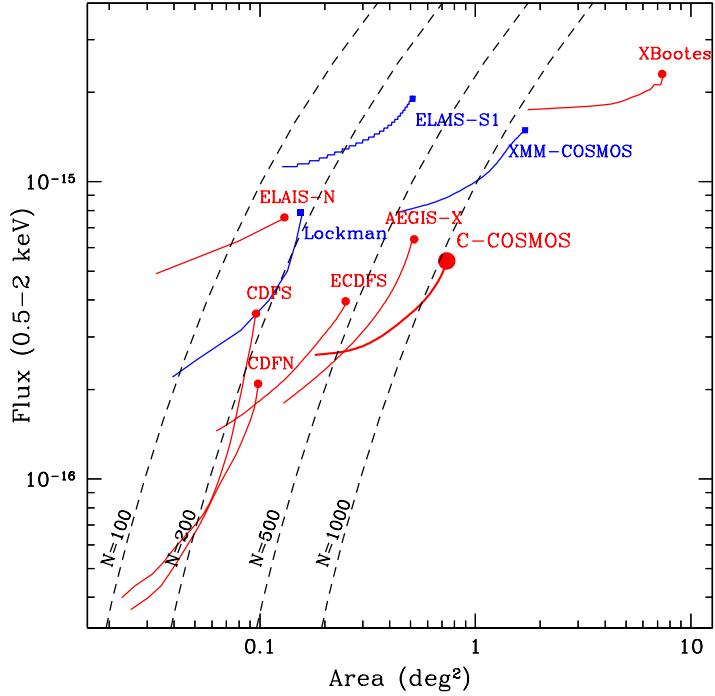}
\end{center}
\vspace{-6mm}
\caption{\small Area-Flux curves for {\em Chandra} (red) and {\em XMM-Newton}
  (blue) contiguous X-ray surveys.  Each survey has been plotted using each
  sensitivity curve starting from the flux corresponding to the area that is
  80\% of the maximum area for that survey (large points at the top of each
  curve), to the flux corresponding to the 20\% of the total area (bottom of
  each curve). Data were taken from the following {\em Chandra} references:
  XBootes - Murray et al. 2005; CDFN - Alexander et al. 2003; CDFS - Luo et
  al. 2008; ECDFS - Lehmer et al. 2005; AEGIS-X - Laird et al. 2008; ELAIS-N -
  Manners et al. 2003; and XMM-Newton references: XMM-COSMOS - Cappelluti et
  al. 2009; Lockman - Brunner et al. 2008; ELAIS-S - Puccetti et al. 2006. The
  black dashed curves show the total number of 0.5-2~keV sources expected based
  on the logN-logS relation predicted by Gilli et al. (2007) at the 80\% area
  point. \normalsize }
\label{AreaFluxPoints}
\end{figure}

Hence, to compare the area and depth of C-COSMOS with comparable contiguous {\em
  Chandra} and {\em XMM-Newton} surveys in a consistent fashion requires a
slightly revised measure of area and depth. We have used the Area-Flux plot from
each survey to derive the flux at the point where each survey reaches 80\% of
the maximum survey area.  We plot these values in Figure~\ref{AreaFluxPoints}
(filled circles) for the {\em Chandra} contiguous area surveys (CDFN, Alexander
et al. 2003; CDFS, Luo et al. 2008; ECDFS, Lehmer et al. 2005; AEGIS- X, Laird
et al. 2008; XBootes, Murray et al. 2005; ELAIS-N, Manners et al. 2003), and for
the {\em XMM-Newton} contiguous surveys that fill regions of the flux-area plane
(ELAIS-S1, Puccetti et al. 2006; XMM-COSMOS, Cappelluti et al. 2009;
Lockman-Hole, Brunner et al. 2008). The C-COSMOS flux at 80\% of the area
covered (0.72 sq.deg) in the Soft band is
$6\times$10$^{-16}$erg~cm$^{-2}$s$^{-1}$.

Compared with other plots of this kind (e.g. Brandt \& Hasinger 2005) survey
points in Figure~\ref{AreaFluxPoints} tend to be moved diagonally toward smaller
area and high flux limits. This shift can be quite large for surveys with
shallow slopes at low fluxes in their area-flux~limit curves (as for examples the deep 
fields). This is because
the normally quoted area is the maximum area of the survey, while the normally
quoted flux limit is that of the faintest source in the survey, which can be
detected only in a much smaller area.

Curves of constant numbers of sources (for the Soft band) are shown in
Fig.~\ref{AreaFluxPoints} following the predictions of Gilli et al. (2007) XRB
model\footnotemark.  
\footnotetext{The curves have been computed using the tool ``POrtable Multi
  Purpose Application for XRB and AGN counts'' available at the web site
  http://www.bo.astro.it/$\sim$gilli/counts.html.}  
The larger numbers in XBootes and the two COSMOS surveys are notable. Some 1000
sources are predicted for C-COSMOS above the '80\% area' flux limit in the Soft
band based on the logN-logS relation of Gilli et al. (2007), while 1023 are
actually detected. This can be compared with the CDF fields which have $\sim$200
sources each.

A single number does not convey the complete picture, of course. We also show in
Fig.\ref{AreaFluxPoints} the Area-Flux curve of each survey down to 20\% of the
area.  These curves better explain the differences between the surveys, notably
between the two CDF deep fields, that are due to the different observation's
strategy [changing only the roll angle (CDF-N) or also moving the centroid
(CDF-S)].  The more sensitive, smaller area, parts of each survey add more
sources than indicated by the dashed black lines, especially for the curves that
are closer to vertical.  For example, the AEGIS-X survey (Laird et al. 2009) has
1032 soft sources, about double the number predicted at the 80\%
point. C-COSMOS, with a flatter flux-area curve, has a total of 1340 S band
sources, $\sim$30\% higher than the 80\% area number.

Each of these surveys has extended multi-wavelength coverage, but C-COSMOS is
the only deep and wide X-ray survey field selected for both existing deep
multi-wavelength coverage, and for future legacy value, due to the equatorial
location of the COSMOS field.  The AEGIS field ($\delta$=+52$^{\circ}$), the
CDF-N field ($\delta$=+62$^{\circ}$) and the XBootes field
($\delta$=+35$^{\circ}$) are all too northerly to be accessible by ALMA or the
VLT. The COSMOS field was also selected to have low IR cirrus emission, and a
lack of bright stars, X-ray or radio sources in the field to maximize
multi-wavelength coverage.

C-COSMOS and XMM-COSMOS complement one another by providing large samples of
sources over a wide flux range (Fig.~\ref{fluxisto}), while sharing the same
extensive multi-wavelength data set. XMM-COSMOS provides a larger sample of
extended sources, while C-COSMOS provides a larger sample of starburst and
normal galaxies.

\section{Observations}

A summary of the {\em Chandra} ACIS-I C-COSMOS observations as carried out is
given in Table~\ref{tab:obs_summary}.  Primarily because of thermal constraints
on spacecraft components (POG, \S 3.3.3), many of the 36 C-COSMOS pointings were
scheduled as two or more separate ObsIDs, giving 49 C-COSMOS observations in
all. The indices X-Y (1-1 through 6-6) describe the field numbers, where X is an
index in RA and Y an index in Dec, with 1-1 being in the top right (NE) corner
of Fig.~\ref{tiling} (left panel), and 1-6 being in the top left (NW) corner.

The observations took place in two main blocks: 2006 December - 2007 Jan and
2007 April-June (Table~\ref{tab:obs_summary}).  The fields were observed at
nominal roll angles of 250/70 deg, where the visibility of the COSMOS field is
at maximum ($\sim$70\%) and the pitch angle is such that the constraints are
either unrestricted or restricted only to avoid overheating of the charged
particle detector (EPHIN).  As an equatorial field, the roll angle of the COSMOS
field is quite stable (55.2--69.6, 248.4--256.2) for periods of $\sim$100 days.
As a result, the {\em Chandra X-ray Center} (CXC) Mission Planning team were
able to maintain a tight roll angle range of $\pm$6$^{\circ}$ around the nominal
values (Table~\ref{tab:obs_summary}), leading to a highly uniform exposure of
the whole field.

The mean effective exposure time per field (not per ObsID) is
46.3~ksec, when only the Good Time Intervals (GTIs), cleaned of the
few high background times (\S 4.2.1) are used\footnotemark.  The
maximum exposure is 48.3~ksec and the minimum exposure (excluding a
single 37.6~ksec exposure for field 2-5, Table~\ref{tab:obs_summary})
is 44.1~ksec.  So, with this exception, the range of exposures over
the fully covered inner region varies by just $\pm$2.0~ksec (4\%). 
\footnotetext{This is $\sim$93\% of the requested exposure, well
within the 90\% tolerance limit defined for {\em Chandra} scheduling.}

\begin{center}
\begin{table}
\scriptsize
\caption{\em{Chandra}-COSMOS observation summary}
\label{tab:obs_summary}
\begin{tabular}{ccccccr}
\tableline
\tableline
         & Field & Obs. Start & Exp. Time\tablenotemark{a} & RA  & Dec & Roll \\
 Obs. ID &       & (UT)       & (ks)          & (J2000.0) & (J2000.0) & (deg) \\
\tableline
 7995 & 1-1 & 2007 Jun 01, 03:41 & 44.6 & 10 02 02.05 & +02 22 36.46 & 248.4 \\
 7996 & 1-2 & 2006 Dec 28, 11:28 & 44.7 & 10 01 31.99 & +02 25 20.48 & 63.4 \\
 7997 & 1-3 & 2006 Dec 30, 21:10 & 44.5 & 10 01 01.92 & +02 28 04.50 & 62.8 \\
 8494 & 1-4 & 2006 Dec 16, 13:21 & 20.2 & 10 00 31.85 & +02 30 48.52 & 66.4 \\
 8122 & 1-4 & 2007 Jan 20, 10:15 & 28.0 & 10 00 31.85 & +02 30 48.52 & 55.2 \\
 8493 & 1-5 & 2006 Dec 12, 18:07 & 19.3 & 10 00 01.79 & +02 33 32.55 & 66.4 \\
 7998 & 1-5 & 2007 Jan 10, 21:41 & 26.9 & 10 00 01.79 & +02 33 32.55 & 63.2 \\
 8478 & 1-6 & 2006 Nov 24, 10:17 & 17.6 & 09 59 31.72 & +02 36 16.58 & 69.6 \\
 7999 & 1-6 & 2006 Nov 25, 09:24 & 29.0 & 09 59 31.72 & +02 36 16.58 & 69.6 \\
 8000 & 2-1 & 2007 May 26, 20:23 & 45.2 & 10 01 51.10 & +02 15 05.52 & 253.2 \\
 8001 & 2-2 & 2007 Apr 02, 03:42 & 47.3 & 10 01 21.03 & +02 17 49.54 & 256.2 \\
 8123 & 2-3 & 2007 Apr 07, 13:40 & 48.3 & 10 00 50.97 & +02 20 33.55 & 255.2 \\
 8002 & 2-4 & 2006 Dec 19, 04:57 & 28.5 & 10 00 20.90 & +02 23 17.58 & 65.0 \\
 8496 & 2-4 & 2006 Dec 23, 12:05 & 17.8 & 10 00 20.90 & +02 23 17.58 & 65.0 \\
 8003 & 2-5 & 2007 Apr 02, 17:53 & 37.6 & 09 59 50.83 & +02 26 01.61 & 255.2 \\
 8004 & 2-6 & 2006 Nov 27, 02:25 & 15.3 & 09 59 20.76 & +02 28 45.64 & 68.6 \\
 8482 & 2-6 & 2006 Dec 02, 09:05 & 10.2 & 09 59 20.76 & +02 28 45.64 & 68.6 \\
 8483 & 2-6 & 2006 Dec 04, 03:02 & 21.3 & 09 59 20.76 & +02 28 45.64 & 68.6 \\
 8005 & 3-1 & 2007 Apr 25, 02:42 & 30.8 & 10 01 40.15 & +02 07 34.57 & 255.2 \\
 8552 & 3-1 & 2007 Apr 26, 09:33 & 14.4 & 10 01 40.15 & +02 07 34.57 & 255.2 \\
 8124 & 3-2 & 2007 Apr 08, 03:42 & 31.1 & 10 01 10.08 & +02 10 18.59 & 255.2 \\
 8549 & 3-2 & 2007 May 05, 17:17 & 17.2 & 10 01 10.08 & +02 10 18.59 & 255.2 \\
 8503 & 3-3 & 2006 Dec 31, 10:18 & 20.0 & 10 00 40.02 & +02 13 02.61 & 62.2 \\
 8006 & 3-3 & 2007 Jan 01, 11:48 & 25.8 & 10 00 40.02 & +02 13 02.61 & 62.2 \\
 8007 & 3-4 & 2006 Dec 19, 22:18 & 21.1 & 10 00 09.95 & +02 15 46.64 & 64.2 \\
 8497 & 3-4 & 2006 Dec 25, 01:50 & 27.1 & 10 00 09.95 & +02 15 46.64 & 64.2 \\
 8008 & 3-5 & 2007 Jan 02, 04:39 & 45.0 & 09 59 39.88 & +02 18 30.67 & 61.9 \\
 8009 & 3-6 & 2007 Jan 02, 18:06 & 44.8 & 09 59 09.81 & +02 21 14.70 & 61.8 \\
 8010 & 4-1 & 2007 Apr 27, 18:45 & 32.9 & 10 01 29.19 & +02 00 03.29 & 255.2 \\
 8553 & 4-1 & 2007 Apr 29, 01:02 & 14.4 & 10 01 29.19 & +02 00 03.29 & 255.2 \\
 8011 & 4-2 & 2007 Apr 04, 04:08 & 45.8 & 10 00 59.13 & +02 02 47.30 & 255.2 \\
 8012 & 4-3 & 2007 Jan 04, 05:30 & 48.0 & 10 00 29.06 & +02 05 31.33 & 61.3 \\
 8013 & 4-4 & 2007 Jan 04, 19:44 & 46.9 & 09 59 58.99 & +02 08 15.36 & 61.1 \\
 8014 & 4-5 & 2007 Jan 05, 09:29 & 44.2 & 09 59 28.92 & +02 10 59.38 & 60.9 \\
 8015 & 4-6 & 2007 Jan 07, 09:53 & 44.1 & 09 58 58.85 & +02 13 43.42 & 60.2 \\
 8550 & 5-1 & 2007 Apr 18, 19:11 & 22.7 & 10 01 18.25 & +01 52 32.34 & 255.2 \\
 8016 & 5-1 & 2007 Apr 19, 20:24 & 23.3 & 10 01 18.25 & +01 52 32.34 & 255.2 \\
 8017 & 5-2 & 2007 Apr 04, 17:55 & 45.3 & 10 00 48.18 & +01 55 16.35 & 255.2 \\
 8018 & 5-3 & 2007 Apr 05, 07:17 & 45.8 & 10 00 18.11 & +01 58 00.38 & 255.2 \\
 8019 & 5-4 & 2007 Apr 06, 23:25 & 48.0 & 09 59 48.04 & +02 00 44.41 & 255.2 \\
 8020 & 5-5 & 2007 Apr 09, 06:12 & 47.8 & 09 59 17.97 & +02 03 28.44 & 255.2 \\
 8021 & 5-6 & 2007 Apr 09, 20:24 & 47.3 & 09 58 47.90 & +02 06 12.48 & 255.2 \\
 8022 & 6-1 & 2007 May 10, 23:28 & 30.9 & 10 01 07.30 & +01 45 01.39 & 251.4 \\
 8555 & 6-1 & 2007 May 12, 16:06 & 16.2 & 10 01 07.30 & +01 45 01.39 & 251.4 \\
 8023 & 6-2 & 2007 Apr 10, 12:49 & 48.3 & 10 00 37.24 & +01 47 45.41 & 255.2 \\
 8024 & 6-3 & 2007 Apr 11, 21:40 & 47.9 & 10 00 07.17 & +01 50 29.44 & 255.2 \\
 8025 & 6-4 & 2007 Apr 12, 11:57 & 47.9 & 09 59 37.10 & +01 53 13.47 & 255.2 \\
 8026 & 6-5 & 2007 Apr 13, 07:31 & 45.8 & 09 59 07.03 & +01 55 57.49 & 255.2 \\
 8027 & 6-6 & 2007 Apr 14, 13:54 & 48.3 & 09 58 36.96 & +01 58 41.53 & 255.2 \\
\hline
\tablenotetext{a}{After GTI and high-background filtering for two
  affected obsids. Intervals of 8.50~ksec and 2.45~ksec (respectively)
  were eliminated from the two affected ObsIDs (8003, 8014). }
\end{tabular}
\end{table}
\end{center}

\section{Data Processing}

The data from the 49 obsids were uniformly processed in two phases
using the CIAO 3.4 software tools\footnotemark
\footnotetext{URLhttp://cxc.harvard.edu/ciao/}
(Fruscione et al. 2006) , the {\em yaxx}\footnotemark
\footnotetext{http://cxc.harvard.edu/contrib/yaxx/} tool and custom
versions of the XMM SAS detect tool {\sc EMLdetect}\footnotemark. 
\footnotetext{http://xmm.esac.esa.int/sas/8.0.0/EMLdetect}
Standard Level-1 and Level-2 processing
pipeline\footnotemark~(ASCDS version 7.6.9) from the CXC were used.
In the first processing phase we determined astrometric corrections
(see below) for each ObsID. These corrections were then applied in the
second phase where we reprocessed all event data starting with Level-1
products.
\footnotetext{Pipeline processing levels are explained at URL:
http://cxc.harvard.edu/ciao/data/sdp.html} 

Data processing involved the following series of steps, as summarized below:
\begin{enumerate}
\item Astrometric corrections ($<$1.1$^{\prime\prime}$) to the
  standard COSMOS frame starting with the CXC supplied standard data
  products (\S\ref{astrometry});

\item Baseline data product creation by re-processing all ObsIDs to a
  standard frame of reference using the new astrometry and standard
  CXC pipelines (\S\ref{baseline});

\item Background reduction using high background time filtering (which
  affects only two ObsIDs) (\S\ref{bkgd});

\item Exposure map creation in the three energy bands F, S, and H,
using the standard CIAO tool sequence (\S\ref{expmap});

\item Calculation of the sky coverage (i.e. the area covered to a
 given flux threshold) in the three energy bands, F, S and H;

\item Candidate source detection using a wavelet technique ({\sc PWDetect}, 
Damiani et al. 1997)\footnotemark;

\item Selection of reliable sources, with a probability of being spurious
  $<$2$\times$10$^{-5}$ in at least one band, using maximum likelihood fitting
  ({\sc EMLdetect}) applied {\em simultaneously to each ObsID} at the positions
  of all candidate sources; Puccetti et al. (Paper~II) shows that {\sc
    EMLdetect} reconstructs the input count rate of simulations well, while both
  {\sc PWDetect} and {\sc detector} underestimate the input count rate by about
  15\%;

\item Reliability checks for all sources using simulations, searches
  for outliers and visual checks (rejected candidate sources were all in the
  wings of bright source PSFs);

\item Aperture photometry of reliable sources. At high fluxes the
    systematic error in the PSF, which is intrinsic to the {\sc EMLdetect}
    method, becomes larger than the statistical error; this systematic error is
    not present for aperture photometry.

\item Derivation of reliability and completeness criteria for the source
    catalog, leading to a logN-logS curve that provides an end-to-end check of
    the source extraction by comparing with other surveys in the same flux range
    (\S\ref{lognlogs}).

\end{enumerate}

Steps 1 to 4 are discussed more fully in the following subsections. Complete
details of the steps from 5 onwards, including details of the simulations and
tests, are given in Paper~II.

\footnotetext{We compared {\sc PWDetect} with the CIAO tool {\sc wavdetect}
  used by most {\em Chandra} deep surveys on a subset of C-COSMOS fields, and
  found no substantive difference in the results; {\sc PWDetect} is a much
  faster algorithm, due to better memory buffering.}

\subsection{Astrometry corrections}
\label{astrometry}

In the first phase we determined accurate astrometric offsets for each ObsID.
The good absolute astrometry produced by {\em Chandra} (0.6\arcsec at 90\%
confidence, POG, \S 5) is still of the order of one ACIS pixel. To avoid a loss
of sensitivity, correcting the astrometry to much less than one pixel error is
needed before merging event files, or stacking.

To this end, we first produced a list of bright X-ray sources for each of the 49
ObsIDs, using the standard CIAO {\sc celldetect} tool. Starting with the
standard ACIS Level-2 data products, we generated a broad-band exposure map for
each ACIS CCD using the CIAO\footnotemark~tools {\sc asphist}, {\sc mkinstmap},
and {\sc mkexpmap}.  These exposure maps and event files were then used as input
to a {\em Chandra}-adapted version of the XMM-SAS tool {\sc EMLdetect} (see next
section), with an input source candidate catalog obtained by running the sliding
cell detection tool {\sc eboxdetect} with a high threshold.  All sources
detected with likelihood parameter $\mathcal{L}>$10 were compared with the CFHT
MegaCam I-band catalog of the COSMOS field (Capak et al.  2007), selecting only
the point-like sources with I magnitudes in the range 18--23.  Using this
restricted magnitude range minimizes systematic effects introduced by bright
stars (saturation) and faint background objects (misidentification), and is
appropriate for sources in this flux range (Brandt \& Hasinger 2005).  An
optical--X-ray position correlation was computed using the likelihood algorithm
included in the SAS task {\sc eposcorr} (Cappelluti et al. 2007, 2009). This
task uses all the possible counterparts of an X-ray source in the field to
determine the most likely coordinate displacement. This method is independent of
the actual spectroscopic identifications, but {\em post facto} all the
identifications have proved to be correct (Paper~III).
\footnotetext{URL: http://cxc.harvard.edu/ciao/}
No statistically significant offset in roll was required for any ObsID, so the
change in roll was set to exactly zero.  The systematic offsets between the
X-ray and the optical positions were always smaller than 1.1~arcsec, with an
average shift of $\Delta$RA= 0.04$^{\prime\prime}$ and
$\Delta$dec=0.25$^{\prime\prime}$.

\begin{figure}[t]
\begin{center}
\includegraphics[width=0.6\textwidth]{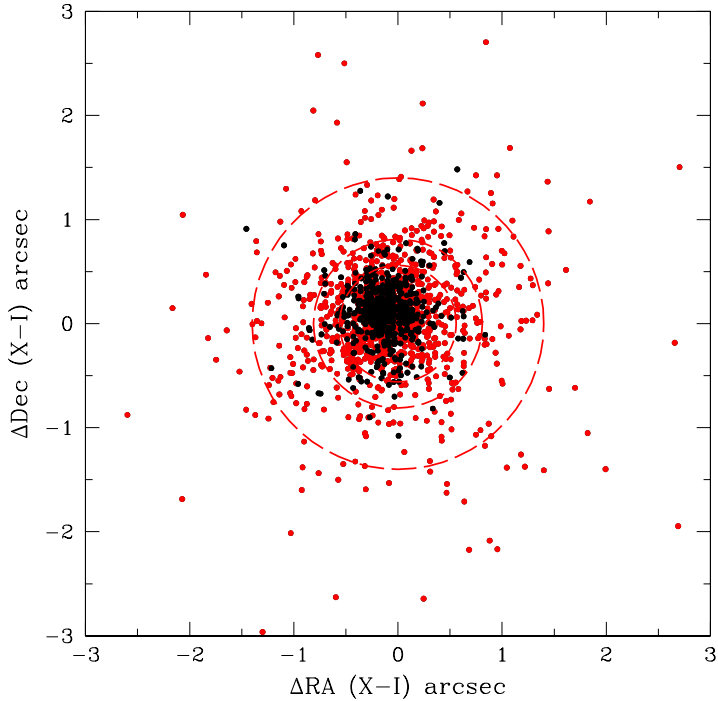}
\end{center}
\vspace{-6mm}
\caption{\small The X-ray to I-band offsets ($\Delta$RA, $\Delta$dec) in arcsec
  for X-ray sources with a secure identification (Civano et al. 2009) after the
  aspect correction described in the text has been applied (\S4.1).  The circles
  encompass 68\% (0.56$^{\prime\prime}$), 90\% (0.81$^{\prime\prime}$) and 95\%
  (1.41$^{\prime\prime}$) of the sources with optical counterparts and secure
  identification.  Red dots mark sources with less than 50 counts in the Full
  band. \normalsize}
\label{astrometry}
\end{figure}

\subsection{Baseline Data Products}
\label{baseline}

The second phase of processing brought the 49 {\em Chandra} ObsIDs to a common
reference frame using the offsets derived above, and generated the baseline data
products that were then used as the starting point in all subsequent C-COSMOS
analysis.

This processing was based on the CIAO thread for
creating a new Level-2 event file from Level-1 products\footnotemark.
\footnotetext{http://cxc.harvard.edu/ciao3.4/threads/createL2/}
First, a new aspect solution for each ObsID was generated to remove the
astrometric offset for each ObsID derived in above section, using the {\sc
  reproject\_aspect}\ tool. Then a new bad pixel file was created using {\sc
  acis\_run\_hotpix} (see 'background reduction' below).  Finally, a new ACIS
Level-2 event file was then created for each ObsID using the {\sc
  acis\_process\_events} tool, with: (a) the standard {\em ASCA} grade set
(grades [0, 2, 3, 4, and 6], POG \S6.14), (b) pixel randomization turned off,
(c) PHA randomization turned on, (d) Very-Faint mode processing enabled, and (e)
the new aspect solution applied.

The astrometric corrections were checked using X-ray sources with point-like
optical counterparts (Civano et al. 2009, Paper~III) that were {\em not} used to
derive the offsets for the individual ObsIDs. The residual systematic shift
(X-ray -- Optical position) is on average $\Delta\alpha$= -0.1$^{\prime\prime}$
and $\Delta$dec=0.08$^{\prime\prime}$, and the 1~$\sigma$ dispersion is
0.56$^{\prime\prime}$ (i.e. the radius within which 68\% of sources lie;
Fig. \ref{astrometry}). We find that 90\% of the X-ray positions agree with the
identified optical/IR counterpart positions to within 1.1\arcsec. The residual
systematic shift is small enough that it will not affect the identification of
any individual source and is smaller than the average X-ray positional error,
and therefore has not been used to correct the astrometry any further. The good
quality of the data provides positions with sub-arcsecond accuracy at off-axis
angle $<$6$\arcmin$, in agreement with other {\em Chandra} surveys
(0.23-1.90\arcsec\ in the CDFS, Luo et al. 2008; 0.3-1.67\arcsec\ in AEGIS,
Laird et al. 2008).

\subsubsection{Background reduction and cosmic ray afterglow detection}
\label{bkgd}

Intervals of high background were determined by creating a background light
curve for the ACIS-I CCD events with point sources found by {\sc wavdetect}\ in
the phase~1 processing removed.  Only two obsids showed intervals with a
significant ($> 5$-$\sigma$) deviation from the quiescent background level (see
Table~\ref{tab:obs_summary}).

Particular care was taken in the rejection of cosmic-ray
afterglows\footnotemark.
\footnotetext{URL: http://cxc.harvard.edu/ciao/why/afterglows.html}
When a cosmic ray hits a CCD pixel a residual charge can remain localized for
tens of seconds and produce ``afterglow events'', that appear to be X-ray
events, at one location for several consecutive CCD frame readouts (POG \S 6.9).
To reject cosmic ray afterglows we used the CIAO tool {\sc
  acis\_run\_hotpix}\footnotemark\ and enabled Very-Faint (VF) mode background
processing in {\sc acis\_process\_events}.  This process was successful as none
of the C-COSMOS sources subsequently detected have the time localization
characteristic of a spurious afterglow source.  This procedure also gave a
25-30\% background reduction in the 0.5-7 keV band.
\footnotetext{http://cxc.harvard.edu/ciao3.4/ahelp/acis\_run\_hotpix.html}

The residual background is very stable over the full field of view at $\sim
1.8\times 10^{-7}$ counts/s/pixel or $\sim$2~counts/200~ksec over a 2 arcsec
radius circle, which represents the typical size of our detection cell across
the field.  Following Alexander et al. (2003), in which the transition between a
photon limited and a background limited regime is defined as $>$3.3 background
counts per detection cell for S/N=3, we conclude that C-COSMOS is photon limited
for point source detection.

\subsubsection{Exposure Maps and Sensitivity Curve}
\label{expmap}

We constructed exposure maps using the standard CIAO tool sequence of
{\sc asphist}, {\sc mkinstmap}, and {\sc mkexpmap}, for
each ObsID on a per-CCD basis, in each of three energy bands, S, H, F.
%

\begin{figure}[t]
\begin{center}
\includegraphics[width=0.6\textwidth]{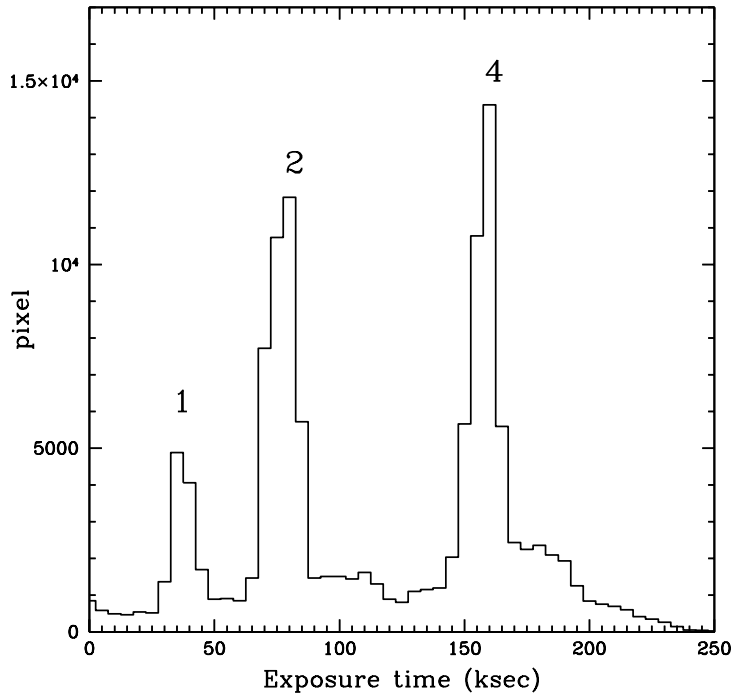}
\end{center}
\caption{\small Histogram of the exposure times in the summed C-COSMOS
field. The narrow peaks lie at the 1, 2 and 4 exposure values. The
broader bases correspond to overlaps caused by slight variations in
the roll angles of the ObsIDs.}
\label{exphisto}
\end{figure}

Figure \ref{tiling} (right panel) shows a composite image of the
effective exposure time (sec) in the Soft band. We clearly see the
central region with four overlapping pointings, the side strips with two
observations, and the corners covered by just one pointing.
The uniformity of the exposure in the central region is shown by the
histogram of the exposure times shown in Fig.\ref{exphisto}. This
histogram shows narrow peaks at the 1, 2 and 4 exposure values, which
have gaussian sigmas of 12.9, 13.6 and 19.3 ksec, respectively, i.e. a
12\% spread on the central region exposure.  The total effective
exposure in the inner, 4 exposure, region is $\sim$160~ksec at the
peak, and $\sim$170~ksec at the mean, in the same region (see
Fig.~\ref{exphisto}).

The C-COSMOS sky coverage (i.e. the area covered as a function of
limiting sensitivity) was computed in the three standard energy bands,
F, S, H using the exposure maps, the background maps and assuming a
spectrum with $\Gamma$ = 1.4 and N$_H$ =N$_H$(Galactic). The sky
coverage in the Soft band is shown in Fig.~\ref{AreaFluxCurve}. 
More details on the Full band and Hard band are given in Paper II (\S 7).

The main uncertainty in the estimated sky coverage comes from the
range of conversion factors from count rates to fluxes induced by the
variety of intrinsic X-ray spectra in the X-ray population, in both
power-law slope and intrinsic absorption, at a minimum. More complex
spectra are surely present. An additional complication is that the
average spectral properties are a function of the observed flux
(Brandt \& Hasinger 2005).
To estimate this uncertainty, we calculated the sky coverage for power
law spectra with $\Gamma=$1.4 and 2.0 with Galactic N$_H$, and for
absorbed power law spectra with $\Gamma=$1.4 and 2.0 and N$_H=10^{22}$
cm$^{-2}$.  The range of conversion factors, given by
PIMMS\footnotemark
\footnotetext{URL: http://http://cxc.harvard.edu/toolkit/pimms.jsp}
is a factor 2.0 in the F band, 1.3 in the H band and 1.2 in the S band
(Table~\ref{conversions}).  As expected from the large width of the Full band,
the uncertainty for the Full band is larger than for the Soft and Hard bands.

\begin{center}
\begin{table}[t]
\footnotesize
\caption{Conversion factors from count rates in the Soft, Full and Hard bands
  (0.5-2, 0.5-7, 2-7 keV) to fluxes in the same bands for different spectral
  assumptions$^a$, computed with the {\em Chandra} Cycle 8 response matrices.}
\smallskip
\begin{center}
\begin{tabular}{ c c c }
\hline
$\Gamma$ & N$_H$     & factor$^b$ \\
         & cm$^{-2}$ &            \\
\hline
\multicolumn{3}{l}{\em Soft Band}\\
1.4 & Galactic & 1.87 \\
1.7 & Galactic & 1.81 \\
2.0 & Galactic & 1.75 \\
1.4 & 10$^{22}$& 2.12 \\
2.0 & 10$^{22}$& 2.15 \\
\hline
\multicolumn{3}{l}{\em Full Band}\\
1.4 & Galactic & 0.75\\	
1.7 & Galactic & 0.89 \\ 
2.0 & Galactic & 1.04 \\
1.4 & 10$^{22}$& 0.51 \\
2.0 & 10$^{22}$& 0.71 \\
\hline
\multicolumn{3}{l}{\em Hard Band}\\
1.4 & Galactic & 0.38 \\
1.7 & Galactic & 0.43 \\
2.0 & Galactic & 0.47\\
1.4 & 10$^{22}$& 0.36 \\
2.0 & 10$^{22}$& 0.45\\
\hline
\end{tabular}

$^a$ $\Gamma$=1.4, N$_H$=Galactic used for catalog fluxes.

$^b$ conversion factor CF where $Flux=B_{rate}/(CF*10^{11})$, in units of
$cts~erg^{-1}~cm^{2}$.
\end{center}
\label{conversions}
\end{table} 
\end{center}

\section{Point Source Catalog}

\subsection{Overview}

In this catalog we report the 1761 sources detected down to a defined threshold
in at least one band. The threshold was chosen to balance completeness (the
fraction of true sources detected) against reliability (the fraction of false
sources detected).  Paper~II describes simulations that allowed us to choose a
threshold which has a known completeness and reliability.  We chose a
probability threshold of P=2$\times$ 10$^{-5}$, giving 99.8\% reliability for
sources with more than 12 counts and 99.7\% reliability for sources with 7
counts. This implies $\sim$3-5 spurious F band detections in the full field with
$>$ 12 counts and 5 spurious detections with $>$ 7.  At this threshold, the
simulations then show that C-COSMOS is 87.5\% complete for 12~count sources and
68\% complete for 7~count sources. The C-COSMOS false source rates are
consistent with those of other surveys (e.g. AEGIS-X, Laird et al. 2008) once
the higher C-COSMOS threshold and larger average source extraction region are
taken into account (see Paper~II, \S 6, 8,9).

The Maximum Likelihood statistic {\em detml} = $-ln(P)$ = 10.8 for
P=2$\times10^{-5}$, and this threshold {\em detml} was applied in {\sc
  EMLdetect}.  The numbers of source detections at or above {\em detml}=10.8 are
listed in the left column of Table \ref{sourcesattwodetml}. Cross-matching the
sources with {\em detml}$>$10.8 in the three bands gives a total of 1761
sources. There are numerous sources with {\em detml}$>$10.8 in fewer than three
bands.  In these cases we can search for significant flux in the other bands to
a 100 times higher P, as the area being searched is now 100 times smaller than
the whole survey area (for a 5\arcsec\ cross-match radius). This corresponds to
a threshold {\em detml} = 6.
In the right hand column of Table \ref{sourcesattwodetml} we give the numbers of
sources detected in each band having 6$<detml<$10.8.
Table~\ref{sourcesinbands} reports the numbers of catalog sources at
or above {\em detml} = 10.8 in 3~bands, 2~bands, or in only one band. (In
this last case the sources must have {\em detml}$>$10.8 in order to
have been selected at all.)

\begin{figure}[t]
\center\includegraphics[width=0.6\textwidth]{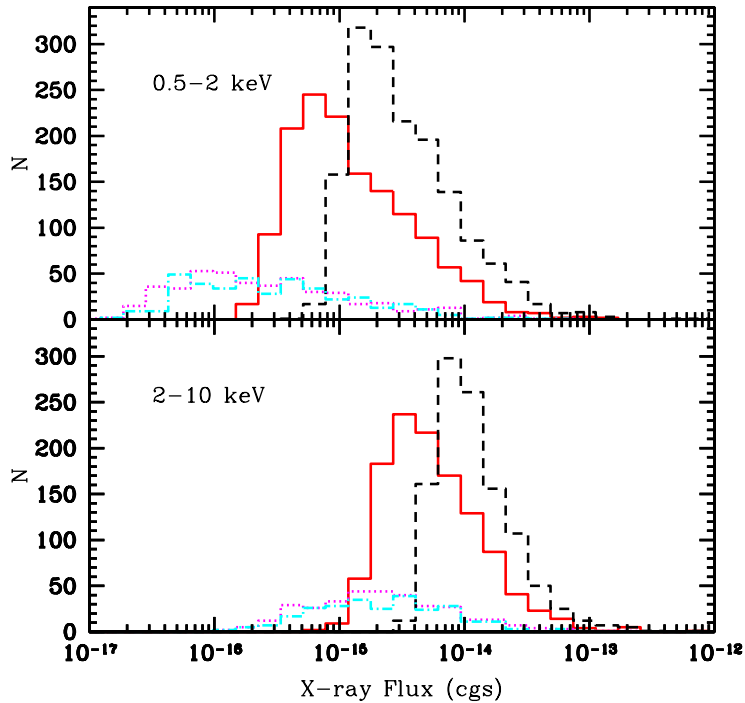}
\caption{\small The distribution of fluxes for sources detected in the
Soft and Hard band (continuous line) compared with the flux
distribution of CDFN sources (dotted line), CDFS (dot-dashed line) and
XMM-COSMOS sources (dashed line). Sources with upper limit have not
been included in this figure. \normalsize}
\label{fluxisto}
\end{figure}

Almost a thousand (946) XMM-COSMOS sources have also been observed by {\em
  Chandra} with an exposure larger than 30 ksec (Cappelluti et al. 2009), and
876 are present in the C-COSMOS catalog. Only 70 sources are not present in the
{\em Chandra} catalog, while 24 XMM-COSMOS sources have been resolved into two
separate sources (Brusa et al. 2009; Paper III) due to the better {\em Chandra}
PSF.  Of the 70 sources not recovered by {\em Chandra}, more than half are in
regions with low exposure (between 30 and 50 ksec) as, for example, in small
gaps of low exposure (Fig.\ref{tiling}). The remainder are either sources with
only hard XMM detections or, after a visual inspection, they are found to be
spurious XMM sources, in agreement with the expected fraction of spurious
sources. C-COSMOS and XMM-COSMOS combine to give a total of $\sim$2800 unique
COSMOS X-ray sources. The distribution of X-ray fluxes for the C-COSMOS sources
in the Soft and Hard bands is shown in Figure \ref{fluxisto}.  For comparison,
we also show the flux distribution of CDFN (dotted line), CDFS (dot-dashed line)
and XMM-COSMOS detected sources (dashed line). The {\em Chandra} and {\em
  XMM-Newton} surveys are complementary in that, together, they span almost 3
orders of magnitude in X-ray flux, and have over 100 soft band (and over 50 hard
band) sources per 0.16 dex bin over about 1.5 orders of magnitude in flux.  The
well-defined cut-off in source numbers at faint fluxes, which reflects the tight
exposure time distribution (Fig.\ref{exphisto}), is significantly different from
the relatively flat distribution of CDFN (dotted line) and CDFS (dot-dashed
line) source fluxes (Fig.\ref{fluxisto}).

\begin{table}[t]
\footnotesize
\caption{Number of sources detected in each band at the two adopted thresholds.}
\begin{center}
\begin{tabular}{l c c }
\hline
Band & $detml\geq$10.8 & 6$<~detml~<$10.8\\
\hline
Full (F)& 1655 & 71 \\
Soft (S)& 1340 & 88   \\
Hard (H)& 1017 &165   \\
\hline
\end{tabular}
\end{center}
\label{sourcesattwodetml}
\end{table} 

\begin{table}[t]
\footnotesize
\caption{Number of sources with $detml\geq$10.8 in at least one band.} 
\begin{center}
\begin{tabular}{lc}
\hline
Bands & Number of sources\\
\hline
F+S+H &~ 922 \\
F+S   &~ 474 \\
F+H   &~ 257 \\
F     &~~ 73 \\
S     &~~ 32 \\
H     &~~~ 3 \\
Total & 1761 \\
\hline
\end{tabular}
\end{center}
\label{sourcesinbands}
\end{table} 

The complete catalog contains source positions and source count rates,
exposure times, signal-to-noise ratio (S/N), and fluxes in the three
bands and hardness ratios (see next section). The catalog is ordered
with the sources detected in the Full band first, followed by those
detected in the Soft band only and by those detected in the Hard band
only.

The resulting catalog is available in the ApJ on-line version and on
the '{\em Chandra} COSMOS Survey' website\footnotemark.
\footnotetext{http://chandracosmos.cfa.harvard.edu/reports/analysis/20090310\_TA\_source\_catalog\_2.1/}
%
Supporting data products (including images, event files and exposure
maps) will be available at the '{\em Chandra} COSMOS Survey' website
and at IRSA\footnotemark.
\footnotetext{URL: http://irsa.ipac.caltech.edu/data/COSMOS/}
At the {\em Chandra} COSMOS Survey it will also be possible to browse
a database that includes 'postage stamps' of the X-ray data for each
source, along with the multiwavelength optical and infrared data,
including the I-band, K-band and Spitzer 3.6$\mu$m (Band~1) images
used in Paper III to identify the sources.

\subsection{Catalog Description}

The {\sc EMLdetect} procedure was run on the three bands: Soft, Hard
and Full.  In order to be consistent with other results in literature,
count rates estimated in the 2--7~keV and 0.5--7~keV energy bands were
extrapolated into 2--10 keV and 0.5--10~keV fluxes, respectively,
using a spectral slope, $\Gamma$=1.4. We also report the number counts
obtained from aperture photometry (see Paper~II).

Table \ref{cata} gives the columns of the catalog of the 1761 X-ray
sources. A more detailed description of each column is reported 
below:

\begin{itemize}

\item[-] {\em Column 1}: Chandra source name, following the standard IAU
  convention with the prefix ``CXOC'' for 'Chandra X-ray Observatory COSMOS'
  survey.

\item[-] {\em Column 2}: Source number. Sources are listed in order of
  detection: first those detected in the Full band with $detml \geq$10.8,
  followed by those detected in the soft band only and by those detected in the
  Hard band only.

\item[-] {\em Column 3-4}: Right Ascension and Declination in the J2000
  coordinate system.

\item[-] {\em Column 5}: Positional error
  ($\sqrt{\sigma_{RA}^2+\sigma_{Dec}^2}$) computed using the following equation
  $Pos_{error}=PSF_{radius}/\sqrt S$ where S is the number of net source counts,
  after the subtraction of the background, in a circular region of radius
  corresponding to the 50\% encircled energy in the field where the source 
  is at the lowest off-axis angle (Paper II).

\item[-] {\em Column 6-7}: Count rate and count rate error in the Full band
  (0.5-7~keV). These are {\em effective} count rates that would apply if the
  source had been observed at the aim point in every pointing. I.e. computed by
  dividing the best fit counts for each source by the effective exposure time at
  the position of each source (the effective exposure time includes corrections
  for vignetting, dither, bad pixels and spatially-dependent quantum
  efficiency).  The count rate error at 68\% confidence level was computed using
  the equation $error=={\sqrt{ C_{s,90\%} + (1+a)B_{90\%}}\over{0.9 \cdot T}}$,
  where where C$_s$ are the source counts estimated by {\it EMLdetect},
  corrected to an area including 90\% of the
  PSF\footnote{http://cxc.harvard.edu/caldb/}, B are the background counts
  evaluated from the background rate (counts/pixel) estimated by {\it EMLdetect}
  multiplied for an area of radius R$_w$, which is the mean of the radii,
  correspondig to 90\% enclosed counts fraction (ECF) of each observation,
  weighted by the observation exposure relative to the total exposure, and T is
  the vignetting corrected exposure time at the position of the source from the
  exposure maps.  We use $a$=0.5, to allow for uncertainties in the background,
  which is computed through the {\sc EMLdetect} procedure (see Paper II for more
  details).

\item[-] {\em Column 8--9}: Full band 0.5--10~keV fluxes and errors were
  computed converting count rates to fluxes using the following formula:
  $Flux=B_{rate}/(CF*10^{11})$, where $B_{rate}$ is the count rate in each band
  as described in column 6, CF is the energy conversion factor 0.742
  $cts~erg^{-1}~cm^{2}$ (and 1.837 and 0.381 for the Soft and Hard, 2--10~keV
  band respectively) appropriate for a power law spectrum with spectral index
  $\Gamma$=1.4 and Galactic column density $N_H=2.7\times 10^{20} cm^{-2}$.  For
  sources not detected in this band, a 90\% upper limit is reported (see Paper
  II for details).

\item[-] {\em Column 10}: Full band signal to noise ratio. 

\item[-] {\em Column 11}: Full band exposure time derived from the
  exposure map. 

\item[-] {\em Column 12--13}: The aperture photometry counts and error in the
  Full band (0.5--7~keV) are derived from event data for each individual Obsid
  and CCD where a source lands. Note that ($F\_rate ~\times~ f\_exptime)~ \neq~
  f\_cts\_ap$. Circular extraction regions corresponding to the 90\% ECF for
  that observation are centered on the source RA, Dec. The individual photometry
  values are then merged to produce a single set of values accounting for the
  ECF for each ObsID, given the different extraction regions needed.

\item[-] {\em Column 14}: Exposure time (ksec) from the same region
  used to generate the aperture photometry.

\item[-] {\em Column 15-23}: Same as columns 6--14 for the Soft band
  (0.5-2 keV).  

\item[-] {\em Column 24-32}: Same as columns 6--14 for the Hard band
  (2-7 keV). Fluxes and errors are computed for the 2-10 keV band with
  the conversion factor quoted above.

\item[-] {\em Column 33-35}: Hardness ratio and 90\% upper and lower
  errors computed as follows: H-S/H+S where H are the counts in the
  Hard band and S the counts in the Soft band. The hardness ratio was
  calculated starting with the {\sc EMLdetect} rate values. Upper and
  lower limits were calculated using the Bayesian Estimation of
  Hardness Ratio code (BEHR, Park et al. 2006).  Pseudo-source and
  background count values were generated using the net count rate,
  background rate (per pixel), and a 3~arcsec source aperture and 5-20
  arcsec aperture for background areas. The aperture photometry was
  unsuitable for this purpose because the individual extraction
  apertures do not have the constant background/source area ratios
  required by the assumptions used in BEHR.

\end{itemize}

\begin{table}
\caption{Data fields in the Catalog.}
\begin{center}
{\small
\begin{tabular}{|rrl|}
\hline
\hline
1 & NAME & Chandra source name \\
2 & Source \# &  source number. \\
3 & RA & Chandra Right Ascension (J2000, hms)\\
4 & DEC & Chandra Declination (J2000, dms)\\
5 & pos\_err & Positional error [arcsec]\\
\hline
6 & f\_rate       & 0.5--7~keV count rate [counts/sec]\\
7 & f\_rate\_err  & 0.5--7~keV count rate error [counts/sec]\\
8 & f\_flux       & 0.5--10~keV Flux [erg~cm$^{-2}$s$^{-1}$]\\
9 & f\_flux\_err  & 0.5--10~keV Flux error [erg~cm$^{-2}$s$^{-1}$]\\
10 & f\_snr       & 0.5--7~keV S/N Ratio\\
11 & f\_exptime   & 0.5--7~keV exposure time [ksec] \\
12 & f\_cts\_ap   & 0.5--7~keV aperture photometry net counts [counts]\\
13 & f\_cts\_ap\_err&0.5--7~keV aperture photometry net counts error [counts]\\
14 & f\_exptime\_ap& 0.5--7~keV exposure time from aperture photometry [ksec] \\
\hline
15& s\_rate       & 0.5--2~keV count rate [counts/sec]\\
16& s\_rate\_err  & 0.5--2~keV count rate error [counts/sec]\\
17& s\_flux       & 0.5--2~keV Flux [erg~cm$^{-2}$s$^{-1}$]\\
18& s\_flux\_err  & 0.5--2~keV Flux error [erg~cm$^{-2}$s$^{-1}$]\\
19 & s\_snr       & 0.5--2~keV S/N Ratio \\
20 & s\_exptime   & 0.5--2~keV exposure time [ksec] \\
21 & s\_cts\_ap   & 0.5--2~keV aperture photometry net counts [counts]\\
22 & s\_cts\_ap\_err&0.5--2~keV aperture photometry net counts error [counts]\\
23 & s\_exptime\_ap& 0.5--2~keV exposure time from aperture photometry [ksec] \\
\hline
24& h\_rate       & 2--7~keV count rate [counts/sec]\\
25& h\_rate\_err  & 2--7~keV count rate error [counts/sec]\\
26& h\_flux       & 2--10~keV Flux [erg~cm$^{-2}$s$^{-1}$]\\
27& h\_flux\_err  & 2--10~keV Flux error [erg~cm$^{-2}$s$^{-1}$]\\
28& h\_snr        & 2--7~keV S/N Ratio \\
29& h\_exptime    & 2--7~keV exposure time [ksec] \\
30& h\_cts\_ap    & 2--7~keV aperture photometry net counts [counts]\\
31& h\_cts\_ap\_err&2--7~keV aperture photometry net counts error [counts]\\
32& h\_exptime\_ap& 2--7~keV exposure time from aperture photometry [ksec] \\
\hline
33 & hr  & hardness ratio \\
34 & hr\_lim\_lo & hardness ratio 90\% lower limit \\
35 & hr\_lim\_hi & hardness ratio 90\% upper limit \\
\hline
\end{tabular}
}
\end{center}
\label{cata}
\end{table}

\subsection{Catalog Completeness \& Number Counts}
\label{lognlogs}

\begin{figure}
\includegraphics[width=0.5\textwidth]{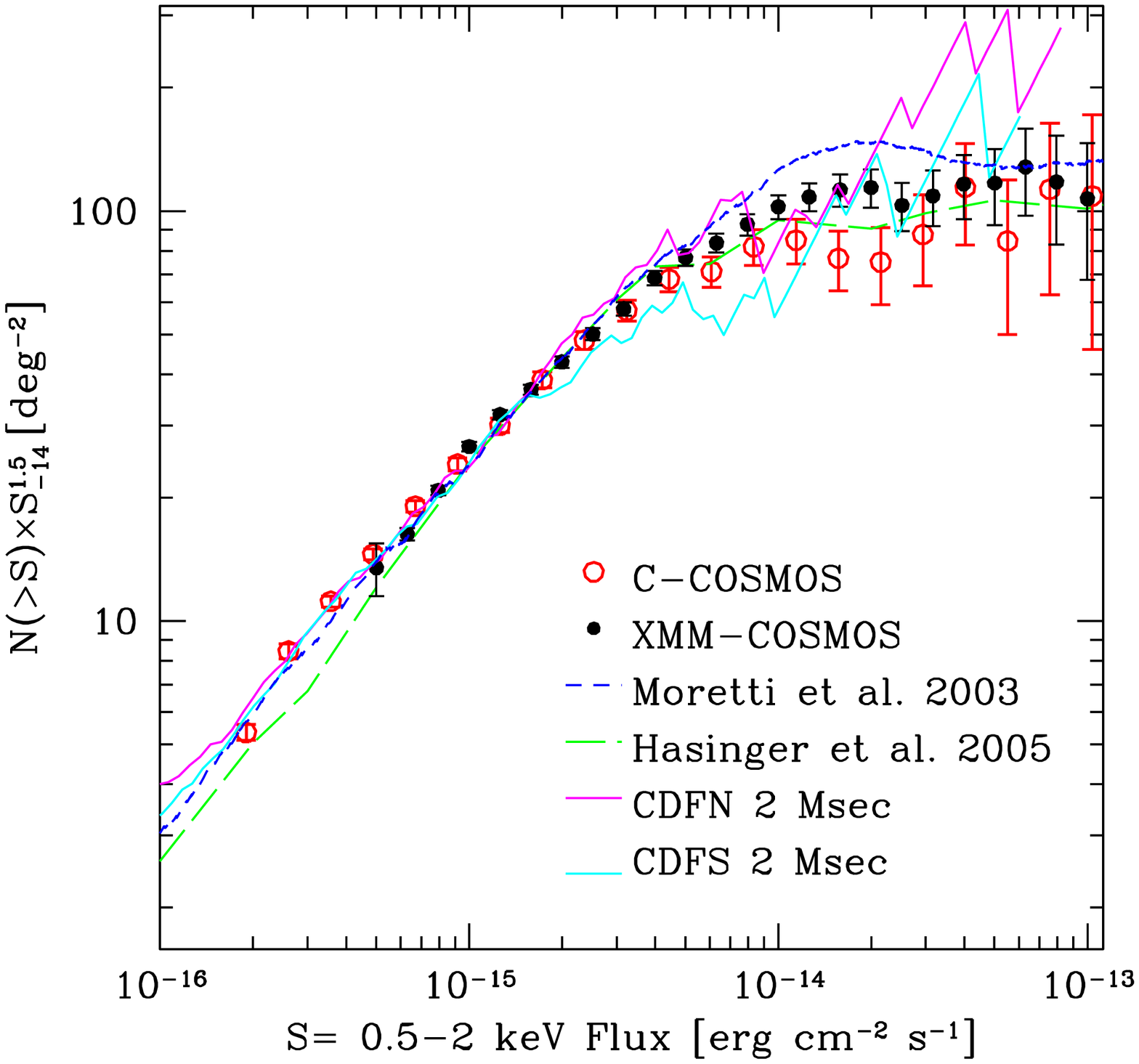}
\includegraphics[width=0.5\textwidth]{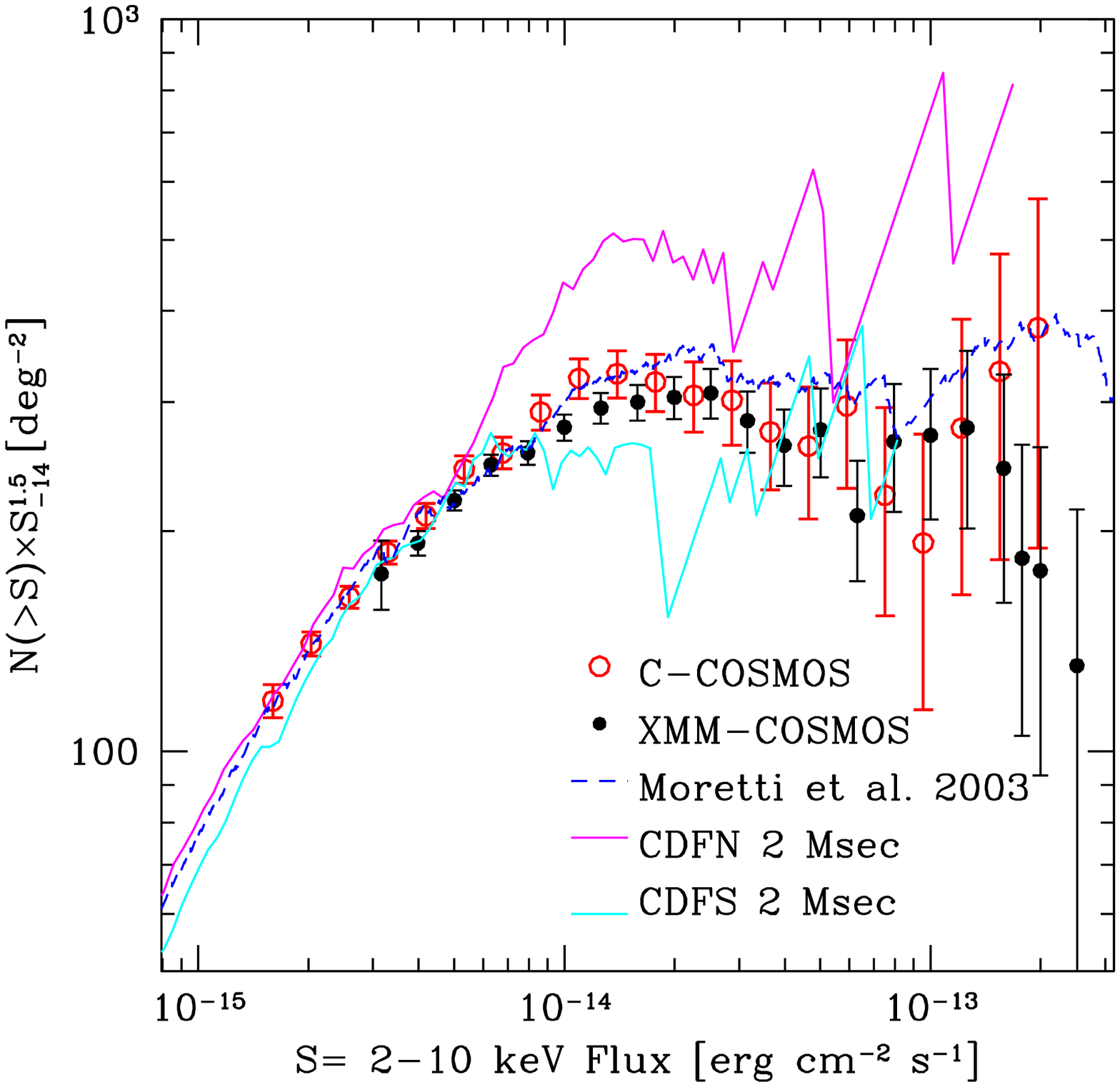}
\vspace{-2.5cm}
\caption{\small The Euclidean-normalized, logN-logS curves for C-COSMOS sources
  with {\em detml}$>$10.8: {\em left:} Soft band (0.5--2~keV, red open circles),
  {\em right:} Hard band (2--10~keV). The XMM-COSMOS curve (black filled
  circles, Cappelluti et al. 2009), the soft band curve of Hasinger et
  al. (2005; green line), the Moretti et al. (2003) compilation (blue dashed
  line), and the CDF-N (magenta solid line, Alexander et al. 2003) and CDF-S
  (cyan solid line, Luo et al. 2008) curves. The agreement is good over the flux
  interval where the various surveys have good statistics (see text).}
\label{lognlogs}
\end{figure}

In order to provide an end-to-end check that the many calibration steps taken in
deriving the {\em Chandra} COSMOS point source catalog have been performed
correctly, we constructed the observed logN-logS curve, i.e. the number of
sources, N($>$S), detected per square degree brighter than a a given flux, S
(erg~cm$^{-2}$s$^{-1}$) in the Soft (0.5-2~keV) and Hard (2--10~keV)
bands. Because at the limiting fluxes the sky coverage is small
(Fig.\ref{AreaFluxCurve}), and so has a large fractional error, we used the flux
limits given in Tab.\ref{tabflux}, column~3, thus omitting the faintest $\sim$10
sources.  X-ray source counts in this flux and energy range have been well
studied, giving us a good baseline against which to compare C-COSMOS (Cappelluti
et al. 2009).

The results, for sources detected at {\em detml}$>$10.8
(Tab.~\ref{sourcesattwodetml}, left column) are shown in figure~\ref{lognlogs},
normalized by a Euclidean 1.5 slope to enable differences between various X-ray
logN-logS curves to be seen easily. Figure~\ref{lognlogs} also shows comparisons
with several other logN-logS curves: from Moretti et al (2003, blue dashed
line), which combines data from ROSAT (for bright sources), XMM-Newton (for
intermediate flux sources), and {\em Chandra} for faint sources; from Hasinger
et al. (2005) logN-logS (green dashed line); and from the CDF-N (magenta solid
line, Alexander et al. 2003) and CDF-S (cyan solid line, Luo et al. 2008)
curves. In the range where these curves overlap and C-COSMOS has good statistics
the agreement is excellent, and C-COSMOS extends a factor $\sim$4 below the
XMM-COSMOS limit, as expected.

In the Soft band, around $\sim$2$\times$10$^{-14}$ erg~cm$^{-2}$s$^{-1}$, the
C-COSMOS logN-logS shows a $\sim$20-30\% underdensity at a 2$\sigma$ level with
respect to the XMM-COSMOS source counts. In order to evaluate this deviation, we
estimated the amplitude of the fluctuations expected due to sample and cosmic
variance. According to Yang et al. (2004, 2006) and Cappelluti et al. (2009),
the fluctuations of the counts in a box of area $\Omega$~deg$^2$ of a population
of $\mathcal{N}$~deg$^{-2}$ sources at a given flux limit, is given by a linear
combination of a Poisson fluctuations and a cosmic variance component introduced
by source clustering:
\begin{equation}
\sigma^{2}_{cv}=\mathcal{N} + \frac{\mathcal{N}}{\Omega^2} \int{w(\theta)d\theta_{1}d\theta_{2}}
%
\label{eq:cv}
\end{equation}
In eq. \ref{eq:cv} 
%
%
$w(\theta$) is the angular autocorrelation function expressed as a
$w(\theta)=\frac{\theta}{\theta_{0}}^{-\gamma}$.  According to
Cappelluti et al. 2007 eq. \ref{eq:cv} can be solved analytically by
knowing the slope and the amplitude of $w(\theta)$.  By using the
source surface density of Soft X-ray sources at 2$\times$10$^{-14}$
erg~cm$^{-2}$s$^{-1}$ (i.e. $\sim$30 source deg$^{-2}$) on a box of
0.9 deg$^{2}$, and assuming the angular autocorrelation function of
Miyaji et al. (2007) for XMM-COSMOS (i.e. $\theta_0$=2\arcsec,
$\gamma$=1.8), we determined $\sigma^{2}\sim$36 which corresponds to a
fraction variance of 20\% of the source counts.  We can therefore
conclude that a deviation of the size observed can be introduced by a
single structure, in an area of XMM-COSMOS not covered by {\em
Chandra}, that generates a fluctuation in the bright source counts at
1.5$\sigma$ level.

Another check of the source detection efficiency at the brighter
C-COSMOS flux levels is a comparison with the XMM COSMOS survey
(Hasinger et al. 2007). As shown by Cappelluti et al. (2009) 
and Brusa et al. (2009, in preparation), C-COSMOS
recovers $\sim$93\% of the XMM sources in the C-COSMOS field,
resolving $\sim$3\% into close pairs.

\section{Conclusions \& Future Work}

We have presented the $\sim$0.9~sq.deg {\em Chandra} COSMOS survey
(C-COSMOS) and a catalog of point sources from that survey. Employing
a heavily overlapping tiling of ACIS-I observations has proven an
effective method of covering a large area to a well-defined exposure
($\pm$12\%) and uniform flux limit. The central $\sim$0.5~sq.deg
achieved an exposure of 160~ksec, and the outer $\sim$0.4~sq.deg
achieved an exposure of $\sim$80~ksec.  The equatorial location of
COSMOS helped to produce a uniform tiling pattern by allowing an
almost constant roll angle for {\em Chandra} observations over most of
the target visibility window.
The point source catalog from the C-COSMOS survey has a flux limit of
2$\times$10$^{-16}$erg~cm$^{-2}$s$^{-1}$ (0.5-2~keV) and contains
1761 sources detected in at least one band with a probability of being
spurious of $<$2$\times$10$^{-5}$ ($detml\geq$10.8).

The novel three-stage source detection method employed (Paper~II) coped well
with the peculiarities of the C-COSMOS tiling scheme and, more generally, is
good at separating close pairs of sources, while retaining photometric
accuracy. The C-COSMOS sky coverage has a sharp cut-off which produces a
homogeneous flux threshold over the whole area and the soft band $logN-logS$
curve for C-COSMOS matches well the Hasinger et al. (2005) determination over a
broad flux range, giving us high confidence in the completeness of the catalog
down to the limiting flux.

The catalog is available in the ApJ on-line version and on
the '{\em Chandra} COSMOS Survey' website (see footnote 16)
%
Supporting data products (including images, event files and exposure
maps) are available at the '{\em Chandra} COSMOS Survey' website
and at IRSA (see footnote 17).
%
%

The sub-arcsecond accuracy of the {\em Chandra} positions, together
with the rich pre-existing deep multiwavelength coverage of the COSMOS
field, allows us to reach a 96\% identification rate for the C-COSMOS
sources with counterparts in both optical and infrared, and 99.7\% in
at least one band (Paper~III).

A parallel effort on the detection of extended sources in the C-COSMOS
field finds $\sim$50 groups and clusters (Finoguenov et al. 2009, in
preparation).

We anticipate a rich haul of science results from C-COSMOS.  The {\em
  Chandra} sources have already resolved ambiguous source
identifications from the XMM-COSMOS survey (Hasinger et al. 2007,
Brusa et al. 2007, 2008, Cappelluti et al. 2007, 2009).  The paper by
Fiore et al. (2008) on the stacking analysis of sources with extreme
mid-infrared to optical ratio, presumably Compton Thick AGN, has been
recently accepted. Several further papers are in preparation or
submitted on: off-nuclear sources in galaxies (Mainieri et al. 2009),
X-ray source correlation functions (Miyaji et al. 2009), the 3D
cluster/AGN cross-correlation function (Cappelluti et al. 2009), high
X-ray/optical flux ratio objects (Civano et al. 2009), high-redshift
QSO (Civano et al. 2009), and other topics.

A basic X-ray spectral analysis of the nearly 500 sources with more
than 80 counts ($\sim$ 23\% of the total sample) becomes possible.
The resulting spectral slopes and absorbing column densities will
allow the statistical properties of a large sample at substantial
redshift and over a uniform and contiguous field to be studied
effectively (Lanzuizi et al. 2009 in preparation).

There is information in C-COSMOS below the current catalog flux limit, thanks to
the low background of {\em Chandra} ACIS.  A 'stacking' analysis (Brusa et
al. 2002, Hornschemeier et al. 2002, 2003) allows the mean X-ray properties of
groups of objects to be determined.  Miyaji et al. (2008) have solved the issues
created by the C-COSMOS tiling scheme for stacking and papers using this tool
are in preparation on z$\sim$1 elliptical galaxies (Kim et al. 2008).
The potential uses of stacking in the C-COSMOS field are extensive,
thanks to the multiple data sets available from which to choose
samples for stacking. For example, there will be $\sim$2$\times$10$^4$
galaxies with good optical spectra from z-COSMOS (Lilly et al. 2007)
in the C-COSMOS field. This entire sample is well characterized both
morphologically via HST imaging, and in terms of stellar population,
from the UV to far-IR coverage of the other telescopes that have
observed COSMOS (Scoville et al. 2007b). This rich data set will
enable galaxy X-ray evolution studies by environment, morphology and
luminosity using fine-grained stacks of C-COSMOS data with $\sim$100
galaxies per bin, for an effective exposure time of $\sim$20~Ms per
bin.

Clearly the C-COSMOS survey will be of value for some time.


\section{Acknowledgements}

We thank the {\em Chandra} mission planning team, especially Pat Slane
and Jan Vrtilek, and the {\em Chandra} pipeline data processing team,
for the extraordinary efforts they put into the successful scheduling
and execution of C-COSMOS.
We also thank the rest of COSMOS team whose support has been
invaluable in reaching this stage.
We thank Bin Luo for sensitivity curves of {\em Chandra} Deep Fields.
We gratefully thank the {\em Chandra} EPO team, in particular Eli
Bressert, for creating the true color X-ray image.
This research has made use of data obtained from the {\em Chandra}
Data Archive and software provided by the {\em Chandra X-ray Center}
(CXC) in the application packages CIAO and Sherpa.

This work was supported in part by NASA {\em Chandra} grant number
GO7-8136A (ME, CV, MB, A.Figonuenov), NASA contract NAS8-39073
(Chandra X-ray Center), and by NASA/ADP grant NNX07AT02G (TM at UCSD).
In Italy this work is supported by ASI/INAF contracts I/023/05/0,
I/024/05/0 and I/088/06, by PRIN/MUR grant 2006-02-5203. In Germany
this project is supported by the Bundesministerium f\"{u}r Bildung und
Forschung/Deutsches Zentrum f\"{u}r Luft und Raumfahrt and the Max
Planck Society.


\end{document}